\documentclass[10pt, twoside]{article}
\usepackage[utf8]{inputenc}
\usepackage[T1]{fontenc}
\usepackage[hypertexnames=false]{hyperref}
\usepackage{bm}

\usepackage[pdftex]{graphicx}
\usepackage{amsmath}
\usepackage{mathrsfs}
\usepackage{amsfonts}
\usepackage{amssymb}
\usepackage{mathtools}
\usepackage{array}
\usepackage{fancybox}
\usepackage{bm}
\usepackage{upgreek}
\usepackage{xcolor}
\usepackage{todonotes}
\usepackage{lineno}
\usepackage{comment}
\usepackage{makecell}
\usepackage{setspace} 
\usepackage{empheq}

\usepackage{pdfpages}
\usepackage{biblatex}
\newcommand{\blat}[1]{\ensuremath{\mathbf{#1}}}
\newcommand{\bgre}[1]{\ensuremath{\bm{#1}}}

\newcommand{\bolda}{\blat{a}}

\newcommand{\boldc}{\blat{c}}

\newcommand{\e}{\blat{e}}

\newcommand{\boldf}{\blat{f}}
\newcommand{\boldF}{\blat{F}}

\newcommand{\boldg}{\blat{g}}

\newcommand{\boldm}{\blat{m}}
\newcommand{\boldM}{\blat{M}}
\newcommand{\n}{\blat{n}}

\newcommand{\boldr}{\blat{r}}

\newcommand{\boldt}{\blat{t}}
\newcommand{\boldu}{\blat{u}}
\newcommand{\boldU}{\blat{U}}

\newcommand{\x}{\blat{x}}

\newcommand{\boldz}{\blat{z}}
\newcommand{\boldZ}{\blat{Z}}

\newcommand{\boldalpha}{\bgre{\upalpha}}
\newcommand{\boldbeta}{\bgre{\upbeta}}
\newcommand{\boldtheta}{\bgre{\uptheta}}

\newcommand{\boldepsilon}{\bgre{\upepsilon}}

\newcommand{\boldchi}{\bgre{\upchi}}

\newcommand{\boldgamma}{\bgre{\upgamma}}
\newcommand{\boldkappa}{\bgre{\upkappa}}
\newcommand{\boldsigma}{\bgre{\upsigma}}

\newcommand{\boldmu}{\bgre{\upmu}}

\newcommand{\lRVE}{\ensuremath{l_0}}
\newcommand{\dd}[1]{\mathrm{\,d}\hspace{0.05em}#1}
\newcommand{\llevicivita}{\ensuremath{\bm{\mathcal{E}}}}
\newcommand{\levicivita}{\ensuremath{\mathcal{E}}}

\newcommand{\mic}{\ensuremath{\mathrm{mic}}}
\newcommand{\mac}{\ensuremath{\mathrm{mac}}}

\newcommand{\intx}{\ensuremath{\mathrm{int}}}
\newcommand{\ext}{\ensuremath{\mathrm{ext}}}

\usepackage[letterpaper,twoside,textwidth=19cm,textheight=25cm]{geometry}

\newcommand{\footremember}[2]{%
	\footnote{#2}
	\newcounter{#1}
	\setcounter{#1}{\value{footnote}}%
}

\begin{document}
\pagestyle{plain}
\pagenumbering{arabic}
	
\title{Macroscopic stress, couple stress and flux tensors derived through energetic equivalence from microscopic continuous and discrete heterogeneous finite representative volumes}
\author{Jan Eliáš\footremember{Brno}{Brno University of Technology, Faculty of Civil Engineering, Brno, Czechia}\footremember{email}{Corresponding author: jan.elias@vut.cz} \and Gianluca Cusatis\footremember{Northwestern}{Northwestern University, Department of Civil and Environmental Engineering, Evanston, IL USA}}
\date{}


\maketitle 
	
\section*{Abstract}
This paper presents a~rigorous derivation of equations to evaluate the macroscopic stress tensor, the couple stress tensor, and the flux vector equivalent to underlying microscopic fields in continuous and discrete heterogeneous systems with independent displacements and rotations. Contrary to the classical asymptotic expansion homogenization, finite size representative volume is considered. First, the macroscopic quantities are derived for a~heterogeneous Cosserat continuum. The resulting continuum equations are discretized to provide macroscopic quantities in discrete heterogeneous systems. Finally, the expressions for discrete system are derived once again, this time  considering the discrete nature directly. 

The formulations are presented in two variants, considering either internal or external forces, couples, and fluxes. The derivation is based on the virtual work equivalence and elucidates the fundamental significance of the couple stress tensor in the context of balance equations and admissible virtual deformation modes. Notably, an~additional term in the couple stress tensor formula emerges, explaining its dependence on the reference system and position of the macroscopic point. The resulting equations are verified by comparing their predictions with known analytical solutions and results of other numerical models under both steady state and transient conditions.

\section*{Keywords}
heterogeneity; mesoscale; discrete model; Cosserat continuum; homogenization; macroscale; energy equivalence; mechanics; Poisson's equation

\section{Introduction}
The aim of this article is to derive macroscopic quantities, specifically the stress tensor ($\boldsigma^{\mac}$), the couple stress tensor ($\boldmu^{\mac}$), and the flux vector ($\bolda^{\mac}$), equivalent to static fields in discrete and continuous heterogeneous systems of finite size. The stress and couple stress tensors emerge from fine scale mechanical formulations with independent displacement and rotations, while the flux vector is associated with various problems described by Poisson's equation, such as heat transfer, electrostatics, mass transport, diffusion, among others. The primary objective of these equations is to provide mean, homogeneous characteristics of microscopic heterogeneous flux fields to assess the macroscopic state of a~heterogeneous system, evaluate its overall loading capacity, or develop phenomenological tensorial constitutive equations.

Numerous materials can be considered heterogeneous at the meso or microscale, and their significance in everyday life motivates the development of complex numerical models capable of predicting their behavior. The most reliable numerical models strive to explicitly and realistically represent material heterogeneities. Such meso or microscale models can be continuous~\parencite{EidSeg-21,ThaEnr-23}, constructed as a~system of interacting discrete particles~\parencite{PanPin-23,KrzNit-23}, or lattice element~\parencite{MasKve-23,SayCha-23,ProBar20}. The desirable robustness and reliability of meso/microscale models, unfortunately, come with intense computational demands. Consequently, these materials are often phenomenologically modeled as a~homogeneous continuum. The connection between these two approaches can be established through computational homogenization by replacing the constitutive routine at each integration point with a~subscale mesoscale model~\parencite{SmiBre-98,FisChe-07,EliCus22,ForPra-01}.

In \emph{discrete} systems, contact forces often involve compressive normal forces and frictional tangential forces within a~vast array of granular materials such as sand, rice, pharmaceutical pills, or powders. However, cohesive interactions can also be present, extending the list to highly heterogeneous materials like concrete, ceramics, or ice. When each particle or particle system is associated with a~physical object of the material's internal structure, the discretization is termed \emph{physical}~\parencite{BolEli-21}. The presence of independent rotations at the microscale, as in discrete mechanical models, leads to an~equivalent homogeneous continuous description known as a~micropolar or Cosserat continuum.

In \emph{continuous} mechanical models, the macroscopic statics and kinematics depend on the formulation at the microscale. The Cauchy formulation with the Boltzmann assumption about the symmetry of the stress tensor lacks independent rotations, resulting in a~macroscale model of Cauchy type. In contrast, the Cosserat formulation at the heterogeneous microscale directly leads to the presence of independent rotations and a~non-symmetric stress tensor at the macroscale, making the macroscale model of Cosserat type.

The evaluation of the stress tensor from discrete systems is attributed to \textcite{Love1927} and \textcite{Web66}; hence, referred to as the Love-Weber formula. It can be expressed using either internal or external forces~\parencite{Cam98,Kry03}. For example \textcite{Bag96} and \textcite{KryRot96} derive both variants. The Love-Weber formula is well-known and has been applied in numerous practical applications. The derivations typically assume steady stateequilibrium, newer works extend it to  dynamics~\parencite{Bag99,ForMil-02,ForMil-03,SaxFor-04,NicHad-13,SmiWen14} or involve hydraulic pressure field~\parencite{GeMan-23}. 
The results are nicely summarized by \textcite{YanReg19}.

Comparatively less literature is available on the derivation of the couple stress tensor. Expressions for the couple stress, similar to the derivations in the present paper, can be found in Refs.~\parencite{ChaLia-90,BarVar01,ChaKuh05}. Especially the work of~\textcite{ChaKuh05} is particularly relevant to the present study as it proceeds with a~similar virtual energetic equivalence. However, it does not discuss the critical point that one needs to constrain the virtual displacements to derive the formulas and that constraint is not unique. 

For problems involving scalar fields, such as heat or electric conduction, diffusion, or mass transport, the equivalent homogeneous continuum is described by the standard Poisson's equation. Macroscopic flux vector expressed with the help of external fluxes can be found in Refs.~\parencite{ZhaZho-11} or \parencite{HadLec18}, while \textcite{EliYin-22} derived it using internal fluxes between particles.

Most homogenization work published in the literature assumes a~vanishing size of the representative volume element (RVE). In this work, instead, the finite size of the representative volume is acknowledged. All macroscopic quantities are derived through energetic equivalence and demonstrate that a~choice regarding the virtual kinematics for the couple stress tensor must be made. Both continuous and discrete systems are considered, and two variants employing external or internal forces are presented. The expression for couple stress is shown to contain an~additional term accounting for the location of the macroscopic reference point. Several examples considering steady-state and transient regimes are included for verification of the resulting equations. The entire derivation is conducted in three-dimensional (3D) space; furthermore the Section~\ref{sec:2D} provides the two-dimensional formulation.

\section{Strong form of the physical problems}
Before starting with the derivations of stress, couple stress, and flux tensors, the relevant equations describing underlying physical problems need to be introduced. The stress and couple stress tensors are associated with the mechanical problem, while the flux tensor is related toPoisson's problem (for example, diffusion, heat transfer, and electrostatics). The mathematical structure of both problems is identical, with three unknown variables being the \emph{primary}, \emph{intermediate}, and \emph{flux} fields. There are three governing equations referred to as the \emph{kinematic}, \emph{constitutive}, and \emph{balance} equations, along with two basic types of boundary conditions known as \emph{kinematic} (or primary, essential) and \emph{static} (or flux, natural). 

\subsection{Short introduction to mechanics of Cosserat continuum}
The \emph{primary} variables in the Cosserat continuum~\cite{Vard18_book} represent kinematic degrees of freedom. These primary variables consist of the standard displacement vector, $\boldu$, which is also found in the Cauchy continuum, and the independent rotation vector, $\boldtheta$. Both $\boldu$ and $\boldtheta$ have three components in three-dimensional space.

The displacementgradient, denoted as $\boldU = \nabla \otimes \boldu = \bm{\upvarepsilon} + \bm{\uppsi}$, can be decomposed into two parts: the symmetric part, $\bm{\upvarepsilon}$, which corresponds to the strain tensor in the Cauchy continuum, and the antisymmetric part, $\bm{\uppsi}$. The rotation vector can also be decomposed into the macrorotation, $\boldtheta_M$, and the microrotation, $\boldtheta_m$. The macrorotation vector is associated with the antisymmetric part of the displacement gradient. The mathematical expressions for the tensors described read
\begin{subequations} \label{eq:cosserat_rotations}
\begin{align}
\bm{\upvarepsilon} &= \frac{1}{2}(\nabla\otimes\boldu+\boldu\otimes\nabla) \label{eq:sympartU}\\
\bm{\uppsi}&=\frac{1}{2}(\nabla\otimes\boldu-\boldu\otimes\nabla) =\llevicivita\cdot\boldtheta_M\\
\boldtheta_m &= \boldtheta - \boldtheta_M 
\end{align}
\end{subequations}
$\llevicivita$ is the Levi-Civita permutation tensor exhibiting total anti-symmetry, $\levicivita_{jkl}=-\levicivita_{jlk}=\levicivita_{ljk}$.

The \emph{intermediate} variables consist of the strain tensor, $\boldgamma$, and the curvature tensor, $\boldkappa$. Both are second-order tensors with 9 components in three dimensions. They are defined by the following \emph{kinematic} equations~\cite{Vard18_book}
\begin{subequations} \label{eq:cosstraincurv}
\begin{align}
\boldgamma &= \bm{\upvarepsilon} - \llevicivita \cdot\boldtheta_m =  \boldU - \llevicivita\cdot\boldtheta  \label{eq:cosstrain}\\
\boldkappa &= \nabla\otimes\boldtheta \label{eq:coscurv}
\end{align}
\end{subequations}

The second governing equation is the \emph{constitutive} equation, which relates strain and curvature to the (generally non-symmetric) stress tensor, $\boldsigma$, and the couple stress tensor, $\boldmu$, serving as \emph{flux} variables. This equation, apart from thermodynamic admissibility, can be defined arbitrarily. 

The final governing equation enforces the \emph{balance} of linear and angular momentum
\begin{subequations} \label{eq:cossbalance}
\begin{align}
-\rho \ddot\boldu + \nabla\cdot\boldsigma + \boldf &= \bm{0} \label{eq:continuum_linear_balance}\\
-\rho J_{\rho} \ddot\boldtheta + \nabla\cdot\boldmu + \llevicivita:\boldsigma + \boldz &= \bm{0} \label{eq:continuum_angular_balance}
\end{align}
\end{subequations}
where $\boldf$ represents an~external volume force, and $\boldz$ represents an~external volume couple, both of which are vectors with three components, $\rho$ is material density and $J_{\rho}$ is the specific moment of inertia (moment of inertia per unit mass~\parencite{Iva23}) with units m$^2$. $J_{\rho}$ is considered a~scalar here but can be a~second order tensor, in general. All governing equations must hold for all points within the domain $\Omega$ with a~volume of $V$ and the boundary $\Gamma$. Using the D'Alembert's principle the inertia terms can be understood as external load varying in time and the whole system can be then treated as steady. This approach will be applied throughout the whole paper, anytime volume force or couple is involved, the inertia terms are assumed to be part of them.

The kinematic boundary conditions prescribe displacements and rotations at the boundary part $\Gamma_u$, while the static boundary conditions express linear and angular balances at the remaining boundary part $\Gamma_t$
\begin{align}
\boldt &= \n\cdot\boldsigma & \boldm &= \n\cdot\boldmu \label{eq:static_bc}
\end{align}
where $\n$ represents the outward normal vector, and the vectors $\boldt$ and $\boldm$ are prescribed tractions and couple tractions, respectively. Note that points on the boundary may have both static and kinematic conditions when applied in different directions. For the sake of simplicity this option is not discussed and it is assumed that $\Gamma_u \cap \Gamma_t = \varnothing$ and $\Gamma_u \cup \Gamma_t = \Gamma$. An~extension to the general formulation is straightforward.

The principle of virtual work, the weak form of the problem, can be established by assuming two arbitrary virtual vector fields, $\delta \boldu$ and $\delta \boldtheta$, both of which are smooth enough (with square integrable derivative) and equal to zero at $\Gamma_u$. Integrating the product of the balance equations~\eqref{eq:cossbalance} and the static boundary conditions~\eqref{eq:static_bc} with these virtual fields, utilizing the divergence theorem, and recognizing that both $\delta \boldu$ and $\delta \boldtheta$ are zero on $\Gamma_u$, one arrives at the principle of virtual work.
\begin{align}
\int_{\Omega} \delta \boldgamma : \boldsigma + \delta \boldkappa : \boldmu \dd{V} &= \int_{\Gamma_t} \delta \boldu \cdot \boldt + \delta \boldtheta \cdot \boldm \dd{\Gamma} + \int_{\Omega} \delta \boldu \cdot \boldf + \delta \boldtheta \cdot \boldz \dd{V} \label{eq:CosseratPVW}
\end{align}
The left-hand side is typically referred to as the virtual work of internal forces and couples, denoted as $\delta W_{\intx}$, while the right-hand side represents the virtual work of external forces and couples, denoted as $\delta W_{\ext}$. Note that this derivation confirms that the $\boldgamma$ and $\boldkappa$ tensors (Eq.~\ref{eq:cosstraincurv}) are energetically conjugate with the stress $\boldsigma$ and couple stress $\boldmu$. One can also define the density of the virtual work of internal actions as $\delta w_{\intx} = \delta \boldgamma : \boldsigma + \delta \boldkappa : \boldmu$, and the \emph{average} density of the virtual works in the domain $\Omega$ as $\delta\bar{w}_{\intx} = \delta W_{\intx}/V$ and $\delta\bar{w}_{\ext} = \delta W_{\ext}/V$.

\subsection{Short introduction to Poisson's problem}
The \emph{primary} unknown field in Poisson's problem is a~scalar potential, and its gradient represents the $\emph{intermediate}$ variable. To simplify matters, terminology similar to that used in mass transport phenomena is adopted. In this context, the primary variable is referred to as pressure, denoted as $p$, the intermediate variable is the pressure gradient, represented as $\boldg$, and the \emph{flux} variable is the flux, indicated by $\bolda$.

The \emph{kinematic} equation that relates pressure and its gradient is expressed as
\begin{align}
\boldg&=\nabla p \label{eq:pressuregrad}
\end{align}  
The \emph{constitutive} equation can take various forms (as long as it is thermodynamically admissible). Its linear versions are known as Darcy's law, Fourier's law, or Ohm's law, depending on the specific problem at hand. The \emph{balance} equation is given by
\begin{align}
c\dot p + \nabla\cdot \bolda - q&=0 \label{eq:massbalancecont}
\end{align}
where $q$ is the source/sink scalar (positive when mass flows inside) and $c$ is the material capacity. As argued in mechanics, the transient term will be understood as a~time dependent source term and therefore will be hidden in the $q$ variable hereinafter. Boundary conditions prescribe the primary variable or the projection of the flux variable. The flux boundary condition ensures flux balance at the boundary part $\Gamma_t$
\begin{align}
j &=\mathbf{n}\cdot\bolda \label{eq:flux_bc}
\end{align}  
with $\n$ being the outward normal, and $j$ is the prescribed normal flux which is positive when directed outside the domain.

The principle of virtual power is derived by integrating the product of the balance equation~\eqref{eq:massbalancecont} and the flux boundary condition~\eqref{eq:flux_bc} with an~arbitrary spatial function of virtual pressure, $\delta p$, which is smooth enough and equals zero on the boundary $\Gamma_u$ where the pressure is prescribed. Applying identical steps as in the case of the mechanical problem, one then arrives at the principal of virtual power
\begin{align}
\int_{\Omega} \delta \boldg \cdot \bolda \dd{V} &= \int_{\Gamma_t} \delta p j\dd{\Gamma} - \int_{\Omega} \delta p  q\dd{V} \label{eq:fluxPVW}
\end{align}
The left-hand side is the virtual power of internal actions, $\delta W_{\intx}$; the right-hand side is a~virtual power of external actions, $\delta W_{\ext}$. The local and average virtual power densities reads $\delta w_{\intx} = \delta \boldg \cdot \bolda$, $\delta\bar{w}_{\intx} = \delta W_{\intx}/V$, and $\delta\bar{w}_{\ext} = \delta W_{\ext}/V$.

Indicial notation will be used along with Einstein summation notation whenever convenient. The components of tensors will be written in light font and their previous index will be moved to superscript, for example $i$th component of normal $\e_N$ is written as $(\e_N)_i = e^N_i$.

\section{Homogenization via principle of virtual work \label{sec:homogenization_continuum}}
The homogeneous continuous material description is employed here to derive expressions for the macroscopic stress tensor, macroscopic couple stress tensor, and macroscopic flux vector in terms of the underlying heterogeneous continuum and a~finite size RVE. For the mechanical problem, it is assumed that the macroscopic continuum is of Cosserat type. Even though such a~choice seems obvious since the discrete model has independent degrees of freedom for rotation, \textcite{ForPra-01} showed that asymptotic expansion homogenization can, under certain assumptions, lead to macroscopic Cauchy continuum. The homogenization from Ref.~\parencite{ForPra-01}, however, studies the asymptotic case of an~infinitely small representative volume element. In our case, we are interested in a~finite volume of discrete model. Moreover, the Cauchy continuum is only a~special case of Cosserat mechanics. If the rotational degrees of freedom at the macroscale coincide with the anti-symmetric components of the displacement gradient (i.e., $\boldtheta_m=\bm{0}$), the derivation would still be valid and the macroscopic continuum would be of Cauchy type.

\subsection{Homogenization of mechanical fields}
\begin{figure}[!b]
\centering \includegraphics[width=13cm]{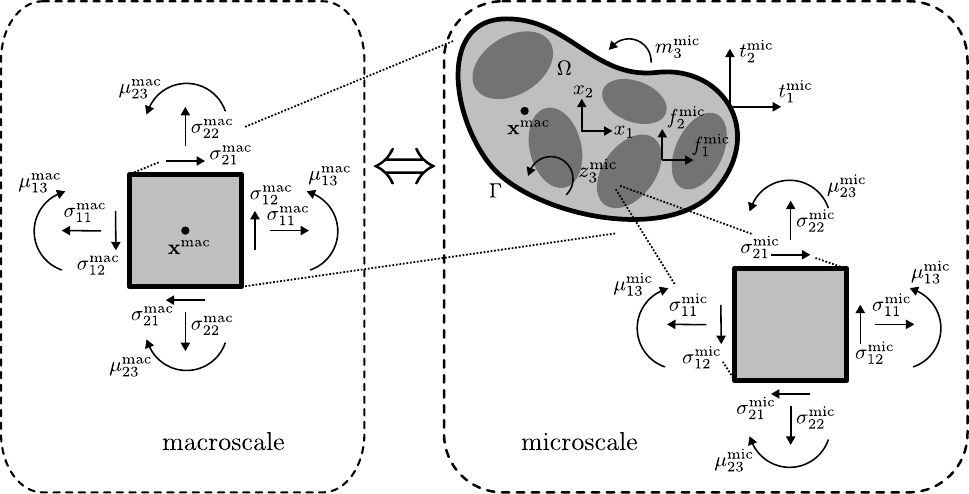}
\caption{Static variables of Cosserat continuum in two dimensions at macro and microscales.}\label{fig:micro_macro_stress}
\end{figure}

Let us assume some balanced body of heterogeneous Cosserat continuum at the microscale in domain $\Omega$  with boundary $\Gamma$ (Fig.~\ref{fig:micro_macro_stress} left-hand side), the internal actions involve stress and couple stress fields are $\boldsigma^{\mic}$ and $\boldmu^{\mic}$, while the external actions are body force $\boldf^{\mic}$, body couple $\boldz^{\mic}$, and tractions and couple tractions at the boundary, $\boldt^{\mic}$ and $\boldm^{\mic}$. Since the body can be subdomain of some large volume, the boundary tractions and couple tractions might expresses the interaction between the domain of interest with the surrounding material. An~equivalent macroscopic stress, $\boldsigma^{\mac},$ and couple stress, $\boldmu^{\mac}$ (Fig.~\ref{fig:micro_macro_stress} right-hand side), will be found by assuming admissible virtual kinematic fields and equating corresponding virtual works at micro and macroscale. The macroscopic stress-like quantities will be then expressed from this equivalence.

Assume that admissible virtual displacements, $\delta \boldu$, and rotations, $\delta \boldtheta$, can be expressed as an~infinite series of terms with gradually increasing power of spatial coordinate \x~\parencite{ChaKuh05,BarVar01}
\begin{align}
\delta u_j(\x) &= \alpha^{(0)}_j + x_i\alpha^{(1)}_{ij} + x_k x_i\alpha^{(2)}_{ikj} + \dots & \delta \theta_j(\x) &= \beta^{(0)}_j + x_i\beta^{(1)}_{ij}  + \dots
\label{eq:expansion_of_displ_and_rot}
\end{align} 
Tensors $\boldalpha$ and $\boldbeta$ contain arbitrary virtual coefficients. The higher power terms are neglected, only the main trends expressed by the initial terms are of interest in this study. Consequently, the virtual works at macro- and micro-scale associated with the higher terms will not be balanced. The rotation expansion contains only two terms, while the expansion of displacements uses three. Other therms could be added, but they would be neglected later because the selected choice of macroscale kinematics and statics cannot accommodate them. 

The above expansions are not admissible unless no virtual displacements and rotations are directly prescribed at any part of the boundary. The kinematic boundary conditions of the domain $\Omega$ are therefore replaced by static boundary conditions with the corresponding tractions, consequently $\Gamma_u\equiv\varnothing$ and $\Gamma_t\equiv\Gamma$. This step does not affect the microscale solution, because the virtual kinematics is not used to solve the microscopic problem. It only allows to virtually vary the kinematics of all the microscopic nodes to account for their static variables at the macroscale. 

By differentiation one computes the virtual strain and curvature tensors
\begin{subequations} \label{eq:expansion_of_strain_and_curv}
\begin{align}
\delta \gamma_{ij} & = \dfrac{\partial \delta u_j}{\partial x_i} - \levicivita_{ijk}\delta \theta_k = \alpha^{(1)}_{ij} + x_k\left(\alpha^{(2)}_{ikj} + \alpha^{(2)}_{kij}\right)  - \levicivita_{ijk}\left(\beta^{(0)}_k + x_l\beta^{(1)}_{lk} \right) + \dots 
\\
\delta \kappa_{ij} &= \frac{\partial \delta \theta_j}{\partial x_i} = \beta^{(1)}_{ij} + \dots 
\end{align} 
\end{subequations}

Equations~\eqref{eq:expansion_of_displ_and_rot} and \eqref{eq:expansion_of_strain_and_curv} are now substituted into the weak form of the balance equation stated as Eq.~\eqref{eq:CosseratPVW}. According to the discussion below the equation~\eqref{eq:CosseratPVW}, the local virtual work density due to internal actions and average virtual work density due to external actions read
\begin{subequations} \label{eq:virtual_works}
\begin{align}
\delta w_{\intx} & = \delta \gamma_{ij} \sigma_{ij} +  \delta \kappa_{ij} \mu_{ij} = \alpha^{(1)}_{ij} \sigma_{ij} + \alpha^{(2)}_{ijk}\left(x_j \sigma_{ik} + x_i \sigma_{jk}\right)  - \beta^{(0)}_i \levicivita_{ijk} \sigma_{jk} - \beta^{(1)}_{ij} \left(\levicivita_{jkl} x_i \sigma_{kl}  - \mu_{ij}\right) + \dots
\label{eq:VW_internal} \\
\delta \bar{w}_{\ext} & = \frac{1}{V}\int_{\Gamma_t} \delta u_i t_i + \delta \theta_i m_i \dd{\Gamma} +  \frac{1}{V}\int_{\Omega} \delta u_i f_i + \delta\theta_i z_i \dd{V} = \frac{1}{V}\int_{\Gamma_t} \alpha^{(0)}_i t_i + \alpha^{(1)}_{ij}x_i t_j + \alpha^{(2)}_{ijk} x_i x_j t_k \label{eq:averageVW_external} \\ &\quad + 
\beta^{(0)}_i m_i + \beta^{(1)}_{ij}x_i m_j  \dd{\Gamma} + 
\frac{1}{V}\int_{\Omega}\alpha^{(0)}_i f_i + \alpha^{(1)}_{ij}x_i f_j + \alpha^{(2)}_{ijk} x_i x_j f_k  + \beta^{(0)}_i z_i + \beta^{(1)}_{ij}x_i z_j \dd{V}  + \dots \nonumber
\end{align} 
\end{subequations}
Notice that the indices have been rearranged so the multiplier of each virtual component is clearly visible. The average virtual work density due to internal actions is obtained by integrating expression~\eqref{eq:VW_internal} over the domain and dividing it by the total volume $V$.  
These equations are used to compute the local virtual work density of internal actions at the macroscale, $\delta w^{\mac}_{\intx}$,  and average virtual work density at the microscale due to internal and external actions, $\delta \bar w^{\mic}_{\intx}$ and $\delta \bar w^{\mic}_{\ext}$. The local version of the macroscale virtual work density reflects that from the viewpoint of the macroscale the domain $\Omega$ is only a~single material point.

The fundamental assumption providing link between the microscale and macroscale imposes equality between the virtual work density of the two scales:
\begin{align}
\delta w^{\mac}_{\intx} &= \delta \bar w^{\mic}_{\intx}  & \delta w^{\mac}_{\intx} &= \delta \bar w^{\mic}_{\ext}  \label{eq:VW_equality}
\end{align} 
where the second equation is derived from the first one through Eq.~\eqref{eq:CosseratPVW}. This fundamental relation, so-called Hill-Mandel macrohomogeneity condition, is often use to create link between microscopic and macroscopic levels, for example in the context of the calculation of macroscopic quantities from discrete systems~\parencite{BarVar01}
and also in FE$^{2}$ homogenization of heterogeneous continuum~\parencite{LarRun-10,KaeRun-21}.

Since the coefficients $\boldalpha$ and $\boldbeta$ can be arbitrary, a~separate equation for each virtual component relating its multipliers must be satisfied. The multipliers can be formally expressed through differentiation of the virtual works with respect to these virtual parameters, equations \eqref{eq:VW_equality} are therefore used in the differentiated form which directly relates the appropriate multipliers, e.g.,  $\partial \delta w^{\mac}_{\intx}/\partial \alpha^{(1)}_{ij} = \partial \delta \bar w^{\mic}_{\intx}/\partial \alpha^{(1)}_{ij}$. The relevant equalities read
\begin{subequations} \label{eq:microbalances}
\begin{align}
\bullet& &&\partial\delta w^{\mac}_{\intx}/\partial\bullet &=&\partial\delta\bar{w}^{\mic}_{\intx}/\partial\bullet &=&\partial\delta\bar{w}^{\mic}_{\ext}/\partial\bullet \nonumber\\
\alpha^{(0)}_{i}&\rightarrow\quad &&0 &=&0 &=& \frac{1}{V}\left(\int_{\Gamma}  t^{\mic}_{i} \dd{\Gamma} + \int_{\Omega} f^{\mic}_i\dd{V}\right) \label{eq:a0}
\\
\alpha^{(1)}_{ij}&\rightarrow\quad && \sigma^{\mac}_{ij} &=&\frac{1}{V}\int_{\Omega} \sigma^{\mic}_{ij} \dd{V} &=& \frac{1}{V}\left(\int_{\Gamma}  x_i t^{\mic}_{j} \dd{\Gamma} + \int_{\Omega}  x_i f^{\mic}_j\dd{V}\right)
\label{eq:a1}
\\
\alpha^{(2)}_{ijk}&\rightarrow\quad &&x^{\mac}_j \sigma^{\mac}_{ik} + x^{\mac}_i \sigma^{\mac}_{jk}  &=&\frac{1}{V}\int_{\Omega} x_j \sigma^{\mic}_{ik} + x_i \sigma^{\mic}_{jk} \dd{V} &=& \frac{1}{V}\left(\int_{\Gamma}  x_i x_j t^{\mic}_{k} \dd{\Gamma} + \int_{\Omega} x_i x_j f^{\mic}_k\dd{V}\right)
\label{eq:a2}
\\
\beta^{(0)}_{i}&\rightarrow\quad &&- \levicivita_{ijk}\sigma^{\mac}_{jk} &=&\frac{1}{V}\int_{\Omega} - \levicivita_{ijk}\sigma^{\mic}_{jk} \dd{V} &=& \frac{1}{V}\left(\int_{\Gamma}  m^{\mic}_{i} \dd{\Gamma} + \int_{\Omega}  z^{\mic}_i \dd{V}\right)
\label{eq:b0}
\\
\beta^{(1)}_{ij}&\rightarrow\quad && \mu_{ij}^{\mac} - \levicivita_{jkl} x^{\mac}_i \sigma^{\mac}_{kl} &=&\frac{1}{V}\int_{\Omega} \mu^{\mic}_{ij} - \levicivita_{jkl} x_i \sigma^{\mic}_{kl} \dd{V} &=& \frac{1}{V}\left(\int_{\Gamma}  x_i m^{\mic}_{j} \dd{\Gamma} + \int_{\Omega}  x_i z^{\mic}_j \dd{V}\right)
\label{eq:b1}
\end{align}
\end{subequations}
At each row all three expressions shall be equivalent. Notice that, based on previous discussion, $\Gamma_t$ is replaced by $\Gamma$. Coordinate  $\x^{\mac}$ refers to the position of the macroscopic point to which energies from the microscale are averaged to.

The microscale variables ($\boldt^{\mic}$, $\boldm^{\mic}$, $\boldf^{\mic}$, and $\boldz^{\mic}$) are known and vary in space. The macroscale variables to be found are the stress tensor and couple stress tensor, $\boldsigma^{\mac}$ and  $\boldmu^{\mac}$. In total there are 18 unknown constants, 9 for $\boldsigma^{\mac}$ and 9 for $\boldmu^{\mac}$. The equality of virtual work requires satisfying 51 equations~\eqref{eq:microbalances} based on 51 virtual parameters $\alpha$ and $\beta$. The first set of equations~\eqref{eq:a0} is satisfied by default, since $\boldalpha^{(0)}$ is associated with rigid-body translation and the microscale domain is assumed to be in a~global equilibrium for which external forces are balanced. The fourth set~\eqref{eq:b0} is essentially only Levi-Civita multiple ($-\llevicivita$) of Eq.~\eqref{eq:a1}, i.e., it is not an~independent equation. One can also argue that it is reasonable to enforce symmetry $\alpha^{(2)}_{ijk}=\alpha^{(2)}_{jik}$ because in Eq.~\eqref{eq:expansion_of_displ_and_rot} they represent identical displacement modes. The number of equations to be satisfied in~\eqref{eq:microbalances} then reduces to 36, which is still too many. Moreover, there is no fundamental reason to exclude additional terms of ansatz~\eqref{eq:expansion_of_displ_and_rot} such as $\boldalpha^{(3)}$, $\boldalpha^{(4)}$, \dots and $\boldbeta^{(2)}$, $\boldbeta^{(3)}$, \dots providing additional infinite number of equations. In other words, the assumed macroscale kinematics does not have sufficiently rich deformation modes to accommodate all the constraints.   Interest is therefore limited only to virtual work associated with shearing and stretching ($\boldalpha^{(1)}$), and bending and torsion (combination of $\boldalpha^{(2)}$ and $\boldbeta^{(1)}$). The remaining equations, including those provided by the higher order virtual displacement and rotation, will remain neglected. 

\subsubsection{Macroscopic stress}

Let us start with Eq.~\eqref{eq:a1}. It corresponds to the following virtual kinematic modes 
\begin{align}
\delta u_j&=x_i \alpha^{(1)}_{ij}  & \delta \theta_j&=0 \label{eq:virtkinematics_stretching}
\end{align}
which clearly express stretching and shearing, see Fig.~\ref{fig:3D_modes} for an~illustration.

The multipliers of $\boldalpha^{(1)}$ in virtual work densities (expressed as derivatives) given by Eq.~\eqref{eq:a1} read
\begin{align}
\frac{\partial \delta w_{\intx}^{\mac}}{\partial \alpha_{ij}^{(1)}} &= \sigma_{ij}^{\mac} &
\frac{\partial \delta \bar{w}_{\intx}^{\mic}}{\partial \alpha_{ij}^{(1)}} &= \frac{1}{V}\int_{\Omega}\sigma_{ij}^{\mic}\dd{V} &
\frac{\partial \delta \bar{w}_{\ext}^{\mic}}{\partial \alpha_{ij}^{(1)}} &= \frac{1}{V}\left[\int_{\Gamma}  x_i t_{j}^{\mic} \dd{\Gamma} +\int_{\Omega}  x_i f_j^{\mic}\dd{V}\right]
\end{align}
One can now equate the derivative of the local and averaged densities of virtual works at micro and macroscale. 
\begin{subequations} \label{eq:macrostress}
\begin{align}
\delta w_{\intx}^{\mac} &= \delta \bar{w}_{\intx}^{\mic} \quad \Rightarrow & \sigma_{ij}^{\mac} &= \left\langle\sigma_{ij}^{\mic}\right\rangle  \label{eq:macrostress_internal}\\
\delta w_{\intx}^{\mac} &= \delta \bar{w}_{\ext}^{\mic} \quad \Rightarrow & \sigma_{ij}^{\mac} &= \frac{1}{V}\left[\int_{\Gamma}  x_i t^{\mic}_{j} \dd{\Gamma} +\int_{\Omega}  x_i f^{\mic}_j\dd{V}\right] \label{eq:macrostress_external}
\end{align}
\end{subequations}
where the operator of the volumetric average of variable $\bullet$ over domain $\Omega$ is used
\begin{align}
\langle \bullet \rangle = \frac{1}{V}\int_{\Omega} \bullet \dd{V} 
\end{align}
Note that this result also satisfies equation~\eqref{eq:b0}, which is just $(-\llevicivita)$ multiple of Eq.~\eqref{eq:macrostress_internal}.

Equation~\eqref{eq:macrostress_internal} gives the macroscopic stress from internal forces at the microscale, while equation~\eqref{eq:macrostress_external} uses only external forces. One can easily transform one to another as demonstrated in Eq.~\eqref{eq:ExtIntstress}. Starting with expression~\eqref{eq:macrostress_external}, using the static boundary condition~\eqref{eq:static_bc}, the divergence theorem, and the balance equation~\eqref{eq:continuum_linear_balance}, one arrives at expression~\eqref{eq:macrostress_internal}: 
\begin{align}
\frac{1}{V}\left[\int_{\Gamma} x_i t^{\mic}_{j} \dd{\Gamma} +\int_{\Omega}  x_i f^{\mic}_j\dd{V}\right] = \left\langle \nabla_k \left(x_i \sigma^{\mic}_{kj} \right)  + x_i f^{\mic}_j\right\rangle =  \left\langle \sigma^{\mic}_{ij} + x_i \left( \nabla_k \sigma^{\mic}_{kj} +  f^{\mic}_j\right)\right\rangle = \left\langle \sigma^{\mic}_{ij}\right\rangle \label{eq:ExtIntstress}
\end{align}

\subsubsection{Macroscopic couple stress \label{sec:couple_stress} }

\begin{figure}[!tb]
\centering \includegraphics[width=\textwidth]{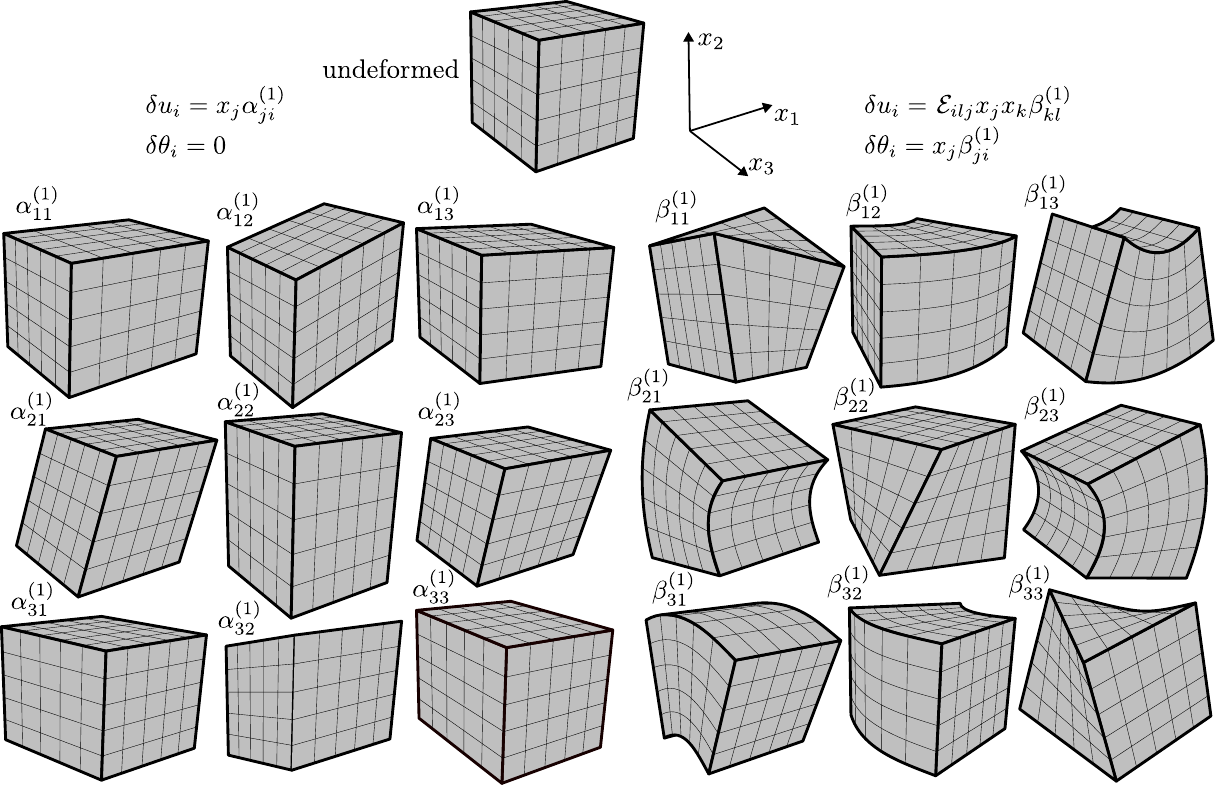}
\caption{Components of displacement fields according to Eqs.~\eqref{eq:virtkinematics_stretching} and \eqref{eq:virtkinematics_bending}: $\alpha$ modes account for stretching and shearing with zero rotations, $\beta$ modes describe curvatures and torsions for which the rotations are not visualized.}\label{fig:3D_modes}
\end{figure}

Next the combination of Eq.~\eqref{eq:a2} multiplied by the Levi-Civita tensor and Eq.~\eqref{eq:b1} is satisfied. The chosen combination can be expressed by relating virtual parameters $\boldalpha^{(2)}$ and $\boldbeta^{(1)}$ by $\alpha_{kil}^{(2)} = \levicivita_{jkl} \beta_{ij}^{(1)}$. Substituting this to the virtual kinematics from Eq.~\eqref{eq:expansion_of_displ_and_rot}, one can easily see that the virtual displacement modes are restricted to 
\begin{align}
\delta u_l & = \levicivita_{jkl} x_k x_i \beta_{ij}^{(1)} & 	\delta \theta_j &=  x_i \beta_{ij}^{(1)} \label{eq:virtkinematics_bending}
\end{align}
expressing bending and torque kinematic modes. Virtual kinematics from both equations \eqref{eq:virtkinematics_stretching} and \eqref{eq:virtkinematics_bending} is visualized in Fig.~\ref{fig:3D_modes}.

The couple stress tensor is going to be derived from the virtual work equality associated with this bending and torque. Eq.~\eqref{eq:b1} is summed with $\levicivita_{jkl}$ multiple of Eq.~\eqref{eq:a2} where the indices $i$, $j$, and $k$ are renamed: $i\rightarrow k$, $j\rightarrow i$, $k\rightarrow l$.
The multipliers of virtual tensors in local and average virtual works at micro and macroscale read (again only terms associated  with arbitrary $\beta^{(1)}_{ij}$ are used)
\begin{subequations} \label{eq:virtdensities_couplestress}
\begin{align}
\frac{\partial \delta w_{\intx}^{\mac}}{\partial \beta_{ij}^{(1)}} &= -\levicivita_{jkl} x^{\mac}_i \sigma^{\mac}_{kl} + \mu^{\mac}_{ij} + \levicivita_{jkl}\left( x_i \sigma^{\mac}_{kl} + x^{\mac}_k \sigma^{\mac}_{il} \right) = \mu^{\mac}_{ij} + \levicivita_{jkl} x^{\mac}_k \sigma^{\mac}_{il} \label{eq:virtdensities_couplestress_a}\\
\frac{\partial \delta \bar{w}_{\intx}^{\mic}}{\partial \beta_{ij}^{(1)}} &=  \left\langle\mu^{\mic}_{ij} + \levicivita_{jkl} x_k \sigma^{\mic}_{il}\right\rangle\\
\frac{\partial \delta \bar{w}_{\ext}^{\mic}}{\partial \beta_{ij}^{(1)}} &=  \frac{1}{V}\left[\int_{\Gamma}  x_i m^{\mic}_{j} \dd{\Gamma} + \int_{\Omega}  x_i z^{\mic}_j \dd{V} + \levicivita_{jkl}\left( \int_{\Gamma}  x_k x_i t^{\mic}_{l} \dd{\Gamma} + \int_{\Omega} x_k x_i f^{\mic}_l\dd{V} \right) \right]
\end{align}
\end{subequations}

Having the derivatives of virtual work densities the equality between them is now enforced and the macroscopic couple stress tensor is derived
\begin{subequations} \label{eq:macrocouplestress}
\begin{align}
\delta w_{\intx}^{\mac} &= \delta \bar{w}_{\intx}^{\mic}\ \Rightarrow & \mu^{\mac}_{ij} &= \left\langle\mu^{\mic}_{ij} + \levicivita_{jkl} x_k \sigma^{\mic}_{il}\right\rangle - \levicivita_{jkl} x^{\mac}_k \sigma^{\mac}_{il} \label{eq:macrocouplestress_internal}\\
\delta w_{\intx}^{\mac} &= \delta \bar{w}_{\ext}^{\mic}\ \Rightarrow & \mu_{ij}^{\mac} &= \frac{1}{V}\left[\int_{\Gamma}  x_i \left(m^{\mic}_{j} + \levicivita_{jkl} x_k t^{\mic}_{l}\right) \dd{\Gamma} + \int_{\Omega}  x_i \left( z^{\mic}_j   + \levicivita_{jkl} x_k f^{\mic}_l\right) \dd{V}\right]  - \levicivita_{jkl} x^{\mac}_k \sigma^{\mac}_{il} \label{eq:macrocouplestress_external}
\end{align}
\end{subequations}
Whenever the reference system origin at the microscale coincides with the location of the macroscopic point, the last term in Eqs.~\eqref{eq:macrocouplestress_internal} and \eqref{eq:macrocouplestress_external} disappears.

Two formulas for the same quantity are obtained, the former one employs internal forces while the latter one uses external forces. Again, it is straightforward to convert one to another using  the static boundary condition~\eqref{eq:static_bc}, the divergence theorem, and the balance equations~\eqref{eq:cossbalance}.
\begin{align}
\frac{1}{V}\left[\int_{\Gamma}\right. & \left.  x_i \left(m^{\mic}_{j} + \levicivita_{jkl} x_k t^{\mic}_{l}\right) \dd{\Gamma} + \int_{\Omega}  x_i \left( z^{\mic}_j   + \levicivita_{jkl} x_k f^{\mic}_l\right) \dd{V}\right] \\=& \left\langle \mu^{\mic}_{ij} + \levicivita_{jkl} x_k \sigma^{\mic}_{il} + x_i \left[\nabla_m \mu^{\mic}_{mj} + \levicivita_{jml} \sigma^{\mic}_{ml} + z^{\mic}_j  + \levicivita_{jkl} x_k \left(\nabla_m\sigma^{\mic}_{ml} + f^{\mic}_l\right)\right] \right\rangle
= \left\langle\mu^{\mic}_{ij} + \levicivita_{jkl} x_k \sigma^{\mic}_{il}\right\rangle  \nonumber
\end{align}

As already pointed out, the couple stress expression is dependent on spatial coordinate $\x$. Let us denote the centroid of the domain $\x_{\mathrm{t}}=\langle \x \rangle$. Equation~\eqref{eq:macrocouplestress_internal} can be rewritten as
\begin{align}
\mu^{\mac}_{ij} = \left\langle\mu^{\mic}_{ij} + \levicivita_{jkl} \left(x_k - x^{\mathrm{t}}_k\right) \sigma^{\mic}_{il}\right\rangle - \levicivita_{jkl}\left( x^{\mac}_k - x^{\mathrm{t}}_k \right)  \sigma^{\mac}_{il}
\end{align}
It means that only the relative position of $\x^{\mac}$ with respect to the centroid of $\Omega$ matters, the absolute position within the reference system is irrelevant. The first term is constant irrespectively of the chosen macroscopic point position. The second term disappears whenever the macroscopic point is placed at the centroid of the domain, which is a~quite natural choice. Otherwise, the macroscopic stress tensor will be actively contributing to the virtual work of bending and torque and therefore affect the couple stress tensor. 

The choice to combine the kinematic modes to equation~\eqref{eq:virtkinematics_bending} was somehow arbitrary. If one had chosen different combinations representing the bending and torque,  different expressions of the couple stress tensor would emerge. The chosen combination is, however, quite natural as is sums the couples caused by microscale stresses and couple stresses. Note that Eqs.~\eqref{eq:a2} and \eqref{eq:b1} were not satisfied separately as well as additional equations arising from possible consideration of higher order virtual kinematic modes by including, e.g., $\boldalpha^{(3)}$, $\boldalpha^{(4)}$ or $\boldbeta^{(2)}$ into Eq.~\eqref{eq:expansion_of_displ_and_rot}. These balances are lost because the assumed macroscale does not have rich enough deformation modes to accommodate them.

Note, that the resulting equation differs from the formula obtained by \textcite{ForPra-01} through asymptotic expansion homogenization, where the term accounting for the effect of microscale stress is missing. Since the homogenization studies the limit for zero size microscale volume, such micro-stresses do not contribute to the macroscale couple stress because they act on infinitely small distance ($\x$ in Eq.~\ref{eq:macrocouplestress}).

\subsection{Homogenization of flux}
The same approach which was used for mechanical fields is applied here to derive the macroscopic flux vector, $\bolda^{\mac}$. The heterogeneous domain at microscale is denoted $\Omega$, the internal action is the the microscopic flux field satisfying the Poisson's equation, $\bolda^{\mic}$, and the external action is volume source $q^{\mic}$ and boundary outward flux $j^{\mic}$.The first step is, similarly to Eq.~\eqref{eq:expansion_of_displ_and_rot}, an~assumption that the admissible virtual primary field can be expressed via arbitrary parameters $\pi^{(0)}$ and $\pi_i^{(1)}$ as 
\begin{align}
\delta p = \pi^{(0)} + x_i \pi_i^{(1)} + \dots \label{eq:virt_press}
\end{align}
The intermediate variable becomes, according to Eq.~\eqref{eq:pressuregrad}, $\delta g_i = \pi^{(1)}_i+\dots$. The weak form of Poisson's problem from Eq.~\eqref{eq:fluxPVW} provides definitions of the densities of virtual powers
\begin{subequations} \label{eq:virtual_powers}
\begin{align}
\delta w_{\intx} & = \delta g_i a_i = \pi^{(1)}_{i} a_{i} + \dots
\label{eq:VP_internal} \\
\delta \bar{w}_{\ext} & = \frac{1}{V}\int_{\Gamma_t} \delta p j^{\mic} \dd{\Gamma} - \frac{1}{V}\int_{\Omega} \delta p q^{\mic} \dd{V} = \frac{1}{V}\int_{\Gamma_t} \pi^{(0)} j^{\mic} + \pi^{(1)}_i x_i j^{\mic}  \dd{\Gamma} -
\frac{1}{V}\int_{\Omega} \pi^{(0)} q^{\mic} + \pi^{(1)}_i x_i q^{\mic}  \dd{V}  + \dots 
\end{align} 
\end{subequations}
Following the homogenization assumption~\eqref{eq:VW_equality} from the mechanical problem, one links the micro and macroscale. Since the virtual parameters $\pi$ are arbitrary, the equality must be valid also for their multipliers obtained by differentiating the virtual power densities with respect to the individual virtual components. The resulting equations read
\begin{subequations} \label{eq:mmicrobalances_flux}
\begin{align}
\bullet& &&\partial\delta w^{\mac}_{\intx}/\partial\bullet &=&\partial\delta\bar{w}^{\mic}_{\intx}/\partial\bullet &=&\partial\delta\bar{w}^{\mic}_{\ext}/\partial\bullet \nonumber\\
\pi^{(0)}&\rightarrow && 0 &=&0 &=& \frac{1}{V}\left(\int_{\Gamma} j^{\mic}\dd{\Gamma} - \int_{\Omega}   q^{\mic}\dd{V}\right) \label{eq:p0}\\
\pi_{i}^{(1)}&\rightarrow && a^{\mac}_i &=& \frac{1}{V}\int_{\Omega} a^{\mic}_i \dd{V} &=&   \frac{1}{V}\left(\int_{\Gamma} x_i j^{\mic}\dd{\Gamma} - \int_{\Omega} x_i q^{\mic}\dd{V}\right) \label{eq:p1}
\end{align}
\end{subequations} 
It was assumed here again that all the boundary belongs to $\Gamma_t$; otherwise the virtual pressure field violates the requirement of being zero on $\Gamma_u$. 

Equation~\eqref{eq:p0} is again satisfied by default, because we assume the microscale is balanced and therefore integral of all external fluxes on the right of line \eqref{eq:p0} must be zero. Equation~\eqref{eq:p1} leads to the definition of macroscopic flux vector
\begin{subequations} \label{eq:macroflux}
\begin{align}
\delta w_{\intx}^{\mac} &= \delta \bar{w}_{\intx}^{\mic} \quad \Rightarrow & a_{i}^{\mac} &= \left\langle a_{i}^{\mic}\right\rangle  \label{eq:macroflux_internal}\\
\delta w_{\intx}^{\mac} &= \delta \bar{w}_{\ext}^{\mic} \quad \Rightarrow & a_{i}^{\mac} &= \frac{1}{V}\left[\int_{\Gamma}  x_i j^{\mic} \dd{\Gamma} - \int_{\Omega}  x_i q^{\mic}\dd{V}\right] \label{eq:macroflux_external}
\end{align}
\end{subequations}
It is again straightforward to convert the expression~\eqref{eq:macroflux_external} using external actions to the one using internal actions~\eqref{eq:macroflux_internal} by Neumann boundary conditions~\eqref{eq:flux_bc}, divergence theorem, and balance equation~\eqref{eq:massbalancecont}.
\begin{align}
\frac{1}{V}\left[\int_{\Gamma}  x_i n_j a_j^{\mic} \dd{\Gamma} -\int_{\Omega}  x_i q^{\mic}\dd{V}\right] = \left\langle  a_i^{\mic} + x_i \left( \nabla_j a_j^{\mic} - q^{\mic}\right) \right\rangle = \left\langle  a_j^{\mic}\right\rangle
\end{align}

As seen in the mechanical problem, the equivalence of virtual works associated with the higher order terms of ansatz~\eqref{eq:virt_press} is neglected. These additional balances cannot be incorporated at the macroscale as no free variables remain there.

\section{Macroscopic quantities derived from discrete systems}
Once the equations for a~heterogeneous continuum are derived, one can proceed towards the homogenization of a~heterogeneous discrete system. There are two distinct options how to evaluate the macroscopic quantities, one can either use (i) external forces, couples, and sources acting on the particles or (ii) internal tractions, couple tractions, and fluxes between them.

\begin{figure}[!b]
\centering \includegraphics[width=15cm]{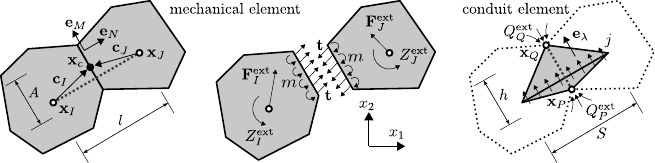}
\caption{Discrete element geometry, internal and external forces, couples and sink/sources in two dimensions.}\label{fig:internal_forces}
\end{figure}

The discrete mechanical system is composed of particles $I$ which behave as rigid bodies with governing nodes $\x_{I}$ where transnational and rotational degrees of freedom are defined. Each particle $I$ is surrounded by other particles $J$ interconnected by mechanical elements $e$ with some area, $A_e$, contact length, $l_e = || \x_J-\x_I||$, and normal direction, $\e_N = (\x_J-\x_I)/l_e$. The internal forces at the contact are given by multiplication of the area by contact tractions, $\boldt_e$, and couple tractions, $\boldm_e$. The same can be written regarding the transport problem, except now conduits elements $d$ connect nodes $P$ and $Q$ and have area $S_d$, length $h_d$ and normal $\e_{\lambda}$. These variables are pictured in Fig.~\ref{fig:internal_forces}.
 
\subsection{Macroscopic quantities evaluated from external actions}

The external actions in discrete system involve forces $\boldF$ and couples $\boldZ$ which account for body load, boundary interaction (for example due to contact with other particle outside the domain $\Omega$), or inertia effects. The continuous forms employing external actions derived in the previous section can be easily transformed into discrete forms.

\subsubsection{Macroscopic stress tensor}
The first integral in Equation~\eqref{eq:macrostress_external} runs over the domain boundary and integrates contributions of external tractions. The second integral brings contributions of volume forces, including inertia. Discretization of Eq.~\eqref{eq:macrostress_external} is done by integrating the external tractions and volume forces into equivalent external forces, $\boldF_f$,  acting at positions $\x_f$, the macroscopic stress in discrete particle or particle system then reads
\begin{align} 
\boldsigma^{\mac} = \frac{1}{V}\sum_f \x_f\otimes\boldF_f \label{eq:fabric_stress_ext_exact}
\end{align}
This expression assumes the true locations of the external forces, $\x_f$. It is often written with respect to individual particles as 
\begin{align}
\boldsigma^{\mac} = \frac{1}{V}\sum_I\sum_{I\!f} \left(\x_I+\boldr_{I\!f}\right)\otimes\boldF_{I\!f} = \frac{1}{V}\sum_I \x_I\otimes \boldF_{I} + \underbrace{\frac{1}{V}\sum_I\sum_{I\!f} \boldr_{I\!f} \otimes\boldF_{I\!f}}_{\text{boundary-radius gap}} \label{eq:fabric_stress_ext_bgap}
\end{align}
where $\boldr_{I\!f}$ is the vector connecting external force location with the particle governing node $\x_I$: $\x_f = \x_I + \boldr_{I\!f}$. The macroscopic stress can be therefore splitted into two contribution, the first one collects the equivalent forces acting at the governing nodes, while the second one accounts for true location of these external forces. According to Refs.~\parencite{Bag96,Bag99,YanReg19} the second term is called \emph{boundary-radius gap}. The name is, however, quite misleading as the difference between external load and governing node locations might occur in the whole volume. For large number of particles the boundary-radius gap becomes negligible and the stress tensor can be estimated only from the first term
\begin{align}
\boldsigma^{\mac} \approx \frac{1}{V}\sum_I \x_I \otimes\boldF_{I} \label{eq:fabric_stress_ext}
\end{align}
The formula without the boundary-radius gap was first presented by~\textcite{DreDeJ72}. The difference between expressions with and without the boundary-radius gap is demonstrated in Sec.~\ref{sec:verif_poisson}. 

\subsubsection{Macroscopic couple stress tensor}
The same process can be used to derive macroscopic couple stress tensor. Starting with definition of the couple stress tensor in the continuum given by Eq.~\eqref{eq:macrocouplestress_external}, the external actions (volume loads, volume couples, tractions and couple tractions) are again discretized into equivalent forces, $\boldF_f$, and couples, $\boldZ_z$, acting at their locations $\x_f$ and $\x_z$.
\begin{align}
\boldmu^{\mac} \approx  \boldsigma^{\mac}\otimes\x^{\mac}:\llevicivita + \frac{1}{V} \sum_z \x_z\otimes \boldZ_z + \frac{1}{V} \sum_f \x_f\otimes \llevicivita:(\x_f\otimes\boldF_f)\label{eq:fabric_couple_stress_ext_exact}
\end{align}
The equation can be also decomposed into a~part collecting equivalent actions at the governing nodes and a~part accounting for the ``true'' position of external load. The second part is referred to as the \emph{boundary-radius gap} based on its similarity with the macroscopic stress decomposition.
\begin{align}
\boldmu^{\mac} &= \boldsigma^{\mac}\otimes\x^{\mac}:\llevicivita +  \frac{1}{V} \sum_I  \x_I \otimes \left(\boldZ_{I} + \llevicivita:\x_I\otimes\boldF_{I}\right)
\label{eq:fabric_couple_stress_ext_bgap}
\\&\quad+ \underbrace{\frac{1}{V} \sum_{I} 
\left[\sum_{I\!z} \boldr_{I\!z} \otimes \boldZ_{I\!z} +  \sum_{I\!f} \left(\x_I \otimes\llevicivita:\boldr_{I\!f} + \boldr_{I\!f} \otimes\llevicivita:\left(\x_I + \boldr_{I\!f} \right)\right)\otimes\boldF_{I\!f}\right]}_{\text{boundary-radius gap}}  \nonumber
\end{align}
The boundary-radius gap can be neglected for large number of particles and the couple stress is then estimated as
\begin{align}
\boldmu^{\mac} \approx  \boldsigma^{\mac}\otimes\x^{\mac}:\llevicivita + \frac{1}{V} \sum_I \x_I\otimes \left(\boldZ_I + \llevicivita:\x_I\otimes\boldF_I\right) 
\label{eq:fabric_couple_stress_ext}
\end{align}
This equation was derived by \textcite{ChaLia-90} and reproduced also in Ref.~\parencite[page 165]{Vard18_book}, in both cases without the first term which disappears when the macroscopic point lies at the origin of the reference system. 

\subsubsection{Macroscopic flux vector}
The macroscopic flux vector, $\bolda^{\mac}$, for discrete Poisson's problems such as mass transport, diffusion, heat conduction or electric flux is derived by discretising  Eq.~\eqref{eq:macroflux_external}. Treating the fluxes at the contact facets as well as source and sink terms (with appropriate signs) as equivalent external fluxes $Q_q$ at locations $\x_q$, the formula can be re-written as
\begin{align}
\bolda^{\mac} = -\frac{1}{V} \sum_q \x_q Q_q \label{eq:fabric_flux_ext_exact}
\end{align}
Transferring the sources to the model governing nodes, it can be restructured as
\begin{align}
\bolda^{\mac} = -\frac{1}{V} \sum_P \x_P Q_P -\underbrace{\frac{1}{V} \sum_P \sum_{P\!q} \boldr_{P\!q} Q_{P\!q}}_{\text{boundary-radius gap}}\label{eq:fabric_flux_ext_bgap}
\end{align}
where $P$ refers to a~discrete unit used as a~control volume for Poisson's problem and $\x_P$ is its governing node bearing the degree of freedom, $\boldr_{P\!q}$ is a~vector from the node $\x_P$ to the ``true'' location of the source term. The control volume $P$ might be both the actual particle as in Ref.~\cite{BolBer04} or a~tetrahedron with the particles at its vertices~\parencite{Gra09,GraBol16}, see Fig.~\ref{fig:model}. To acknowledge the second option, different geometrical variables are used for the discrete Poisson's problem. Here it is only node index, $P$, but the next section adopts dual conduit elements. 

The name \emph{boundary-radius gap} is used for the third time to label the correction term, which can be neglected for large systems with many particles. Then the flux vector becomes
\begin{align}
\bolda^{\mac} \approx -\frac{1}{V} \sum_P \x_P Q_P \label{eq:fabric_flux_ext}
\end{align}
These results has been previously presented in a~similar form by \textcite{HadLec18,ZhaZho-11}.

Equation~\eqref{eq:fabric_flux_ext} violates the first condition from Ref.~\parencite{ChaKuh05} imposed on flux measure: it must be independent of the choice of particle reference points. The original version, Eq.~\eqref{eq:fabric_flux_ext_exact}, is completely independent of the choice of those reference points. The same is valid for simplified Eqs.~\eqref{eq:fabric_stress_ext} and \eqref{eq:fabric_couple_stress_ext} and their full forms~\eqref{eq:fabric_stress_ext_exact} and \eqref{eq:fabric_couple_stress_ext_exact}.

Notice also that equations for stress (\ref{eq:fabric_stress_ext} and \ref{eq:fabric_stress_ext_exact}) and flux (\ref{eq:fabric_flux_ext} and \ref{eq:fabric_flux_ext_exact}) do not depend on the location of the origin of reference system. Moving the origin by vector $\x_0$ adds to the expressions only $\x_0$ multiple of all external sources, which must be zero since the system is in global balance. This is true also for the couple stress equations (\ref{eq:fabric_couple_stress_ext} and \ref{eq:fabric_couple_stress_ext_exact}) providing the first term with macroscopic stress tensor is included.

Finally, notice the negative sign in equations for flux vector. This is due to the chosen convention of source term, which is positive when potential flows in.

\subsection{Macroscopic quantities evaluated from internal actions}
Equations \eqref{eq:fabric_stress_ext_exact}, \eqref{eq:fabric_couple_stress_ext_exact}, and \eqref{eq:fabric_flux_ext_exact} use external actions to express macroscale characteristics. It might be, in some situations, beneficial to work with expressions using internal forces, couples and fluxes. This is, however, possible only under assumption that all external  actions are located at governing nodes of the model, in other words one can express by internal actions only simplified Eqs.~\eqref{eq:fabric_stress_ext}, \eqref{eq:fabric_couple_stress_ext}, and \eqref{eq:fabric_flux_ext} without the boundary-radius gap, not the full equations. On the other hand, the boundary-radius gap terms can be always added when considered to be important.

The balance of internal and external actions must be satisfied for each particle in terms of forces, couples and fluxes.
\begin{subequations} \label{eq:discrete_balance}
\begin{align}
\boldF_I &= -\sum_{e\in I} A_e \boldt_e \label{eq:balance_linear_mom_discrete}\\
\boldZ_I &=  -\sum_{e\in I} A_e\left[\boldm_e + \llevicivita:(\boldc_{Ie}\otimes\boldt_e) \right] \label{eq:balance_angular_mom_discrete} \\ 
Q_P &= \sum_{d\in P} S_d j_d \label{eq:balance_flux_discrete}
\end{align}
\end{subequations}
The vector $\boldc_{I}$ points from $\x_I$ to the centroid of the element face, as shown in Fig.~\ref{fig:internal_forces}. 

\subsubsection{Macroscopic stress tensor}
Substituting the linear momentum balance equation~\eqref{eq:balance_linear_mom_discrete} into the expression for macroscopic stress tensor~\eqref{eq:fabric_stress_ext} yields
\begin{align}
\boldsigma^{\mac} \approx -\frac{1}{V}\sum_I \sum_{e\in I} A_e \x_I \otimes \boldt_e
\end{align}
Every element $e$ connecting nodes $I$ and $J$ is visited twice, once from node $I$ and once from node $J$. The area $A$ is identical but the traction is opposite: $\boldt_{I\!J} = -\boldt_{J\!I}$. The expression can be therefore rewritten into a~single summation over all elements $e$ connecting nodes $I$ and $J$. 
\begin{align}
\boldsigma^{\mac} \approx \frac{1}{V}\sum_e A_e \left(\x_J - \x_I\right) \otimes \boldt_e 
\end{align}
Replacing the difference between coordinates by normal vector multiplied by length gives
\begin{align}
\boldsigma^{\mac} \approx \frac{1}{V}\sum_e A_e l_e \e_{Ne} \otimes \boldt_e \label{eq:fabric_stress_int}
\end{align}
This famous expression is usually attributed to~\textcite{Love1927} and~\textcite{Web66} and it is known under the name Love-Weber formula. It is a~discrete counterpart of Eq.~\eqref{eq:macrostress_internal}. It can be found in many other references, e.g., ~\parencite{RotSel81,ChrMeh-81}. It also exactly corresponds to the formula derived in Ref.~\parencite[Eq.~22]{RezCus16} and Ref.~\parencite[Eq.~31b]{EliCus22} by asymptotic expansion homogenization. Since the period unit cell used in asymptotic homogenization does not have any external actions, the boundary-radius gap is zero and the formula provides exact stress tensor. This is advantageously employed in FE$^2$ homogenization, both for Poisson's and mechanical problems~\parencite{RezCus16,EliYin-22,EliCus22}.

Some publications derive this equations flipped, using $\boldt_e \otimes \e_{Ne}$ instead \parencite{BalMar07,NicHad-13}. The resulting stress vector is then transposed. The mistake seems to be attributed to incorrect balance equation where divergence of stress tensor is taken from the other side: $\boldsigma \cdot \nabla$ instead of $ \nabla \cdot \boldsigma$ as used in Eq.~\eqref{eq:continuum_linear_balance}. Typically this does not really matter as the stress tensor should be symmetric in absence of external or internal couples~\parencite{NicHad-13,LinWu16}. For discrete systems loaded by couples, having angular inertia, or couple tractions at the contacts between particles, the difference can be critical.

\subsubsection{Macroscopic couple stress tensor}
The same approach is taken for the couple stress tensor. Combining the balance of linear~\eqref{eq:balance_linear_mom_discrete} and angular~\eqref{eq:balance_angular_mom_discrete} momentum for each particle with Eq.~\eqref{eq:fabric_couple_stress_ext} gives
\begin{align}
\boldmu^{\mac} \approx \boldsigma^{\mac}\otimes\x^{\mac}:\llevicivita - \frac{1}{V} \sum_I \sum_{e\in I} A_e \x_I\otimes \left[\boldm_e + \llevicivita:((\x_I+\boldc_{Ie})\otimes\boldt_e)\right]
\end{align}
The second position vector $\x_I$ is summed with the arm vector $\boldc_{I}$. The summation results in position of the contact face centroid, $\x_c = \x_I+\boldc_{I}$, see Fig.~\ref{fig:internal_forces}. The identical elements visited twice have opposite tractions and couple tractions, therefore the first appearance of vector $\x_I$ is transformed to $-\e_N l$ as in the previous case of the stress tensor. The couple stress tensor based on internal forces then reads  
\begin{align}
\boldmu^{\mac} \approx\boldsigma^{\mac}\otimes\x^{\mac} :\llevicivita + \frac{1}{V} \sum_e A_e l_e \e_{Ne}\otimes \left[\boldm_e + \llevicivita:(\x_{ce}\otimes\boldt_e)\right] \label{eq:fabric_couple_stress_int}
\end{align}
The equation above is the same as the expression derived in Refs.~\parencite[Eq.~24]{RezCus16} and \parencite[Eq.~31c]{EliCus22} by the asymptotic expansion homogenization. It corresponds to the continuum version from Eq.~\eqref{eq:macrocouplestress_internal}.

\subsubsection{Macroscopic flux vector}
Finally, the macroscopic flux vector expressed by internal forces is derived by combining Eqs.~\eqref{eq:balance_flux_discrete} and \eqref{eq:fabric_flux_ext}
\begin{align}
\bolda^{\mac} \approx -\frac{1}{V} \sum_P \sum_{d\in P} S_d \x_P j_d
\end{align}
The same logic as before transforms the double summation to a~single summation over all elements. The opposite sign of the flux $j$ in two appearances of the same element transforms the $\x_P$ coordinate to $-\e_{\lambda} h$
\begin{align}
\bolda^{\mac} \approx \frac{1}{V} \sum_d S_d h_d \e_{\lambda d} j_d \label{eq:fabric_flux_int}
\end{align}
This result exactly corresponds to the formula derived in Ref.~\parencite[Eq.~35]{EliYin-22} and Ref.~\parencite[Eq.~31a]{EliCus22} by asymptotic expansion homogenization. Its continuous version is equation~\eqref{eq:macroflux_internal}. 

Notice once again that all three equations~\eqref{eq:fabric_stress_int}, \eqref{eq:fabric_couple_stress_int}, and \eqref{eq:fabric_flux_int} were derived by neglecting the boundary-radius gap, therefore, they are not exact and shall be used only for large number of particles.

\section{Virtual work equivalence applied directly to the discrete system}
One can also apply the virtual work concept directly to a~balanced discrete system and derive the equations without going first through the microscale Cosserat continuum. However, displacement and pressure is continuous at the macroscale, but defined at discrete nodes only at the microscale. One therefore cannot incorporate external actions acting outside the nodes into the virtual work equivalence and has to replace them by equivalent actions at the nodes. This exactly correspond to neglecting the boundary-radius gap. All the equations obtained in this section will therefore omit the boundary-radius gap, which can be, however, added if necessary.

Let us start with the mechanical problem. The virtual mechanical work associated with discrete contact $e$ between nodes $I$ and $J$ reads
\begin{align}
\delta W_{\intx}^e &= A l \left( t_i \delta \epsilon_i + m_i \delta \chi_i \right) \label{eq:Piint_singleelem}
\end{align}
The components of virtual strain and curvature of element connecting nodes $I$ and $J$ in the global coordinate system are computed from kinematic equations of the discrete system
\begin{align}
\delta \epsilon_i &= \frac{1}{l}\left[\delta u^J_i-\delta u^I_i+\levicivita_{ijk}\left(\delta\theta^J_j c^J_k - \delta\theta^I_j c^I_k\right) \right] &
\delta\chi_j &= \frac{1}{l}\left(\delta\theta^J_j - \delta\theta^I_j \right) \label{eq:virtual_discrete_kinematic_equation}
\end{align}
Note that the intermediate variables of strain ($\boldepsilon$) and curvature ($\boldchi$) have the same name as in the case of continuum ($\boldgamma$, $\boldkappa$) but different symbols. These are vectors in the discrete case but second order tensors in the continuum.

The restricted virtual kinematics derived by summing equations~ \eqref{eq:virtkinematics_stretching} and \eqref{eq:virtkinematics_bending}, which are visualized in Fig.~\ref{fig:3D_modes}, can be directly applied, yielding 
\begin{align}
\delta u_l &= x_i \alpha^{(1)}_{il} + \levicivita_{jkl} x_k x_i \beta_{ij}^{(1)} & \delta\theta_j &= x_i \beta_{ij}^{(1)} \label{eq:restricted_virt_kinematics}
\end{align}
Of course the previous equation uses relation between $\boldalpha^{(2)}$ and $\boldbeta^{(1)}$ assumed at the beginning of Sec.~\ref{sec:couple_stress}; one can avoid it by using Eq.~\eqref{eq:expansion_of_displ_and_rot} instead. However the same reasoning as provided in Sec.~\ref{sec:homogenization_continuum} will lead to the same result anyway. The strains and curvatures in continuum and discrete systems are given by kinematic equations~\eqref{eq:virtual_discrete_kinematic_equation} and \eqref{eq:cosstraincurv}. To simplify the discrete versions, the following identities are used: $\x_J -\x_I=\e_N l$ and $\x_I + \boldc_I = \x_J + \boldc_J  = \x_c$ (see Fig.~\ref{fig:internal_forces})
\begin{subequations}
\begin{align}
\delta \epsilon_{l} &= e^N_i\alpha_{il}^{(1)} + \levicivita_{jkl}\beta_{ij}^{(1)} x^c_k e^N_i & \delta\chi_j &= e^N_i \beta_{ij}^{(1)} \label{eq:discrete_straincurv}\\
\delta \gamma_{il} &= \alpha^{(1)}_{il} + \levicivita_{jkl} x_k \beta_{ij}^{(1)} & \delta\kappa_{ij} &= \beta_{ij}^{(1)} \label{eq:cont_straincurv}
\end{align}
\end{subequations}

The average density of virtual internal work is obtained by the summation of Eq.~\eqref{eq:Piint_singleelem} over all elements in the system divided by volume. With the virtual strain and curvatures taken from Eq.~\eqref{eq:discrete_straincurv} it reads
\begin{align}
\delta \bar{w}_{\intx}^{\mic} = \frac{1}{V}\sum_e A_e l_e t^e_l e^{Ne}_i \left(\alpha^{(1)}_{il} + \levicivita_{jkl}\beta^{(1)}_{ij}x^{ce}_k\right)  + A_e l_e m^e_j e^{Ne}_i\beta_{ij}^{(1)}
\end{align}
The average density of virtual external work (assuming all forces and couples act at the particle governing nodes) under virtual displacements and rotations from Eq.~\eqref{eq:restricted_virt_kinematics} reads
\begin{align}
\delta \bar{w}_{\ext}^{\mic} = \frac{1}{V}\sum_I  F^I_l \delta u^I_l + Z^I_j \delta \theta^I_j = \frac{1}{V}\sum_I  F^I_l \left( x^I_i \alpha^{(1)}_{il} + \levicivita_{jkl} x^I_k x^I_i \beta_{ij}^{(1)} \right) + Z^I_j x^I_i \beta_{ij}^{(1)}
\end{align}
The density of  the virtual internal work at the macroscale under the restricted virtual kinematics becomes
\begin{align}
\delta w_{\intx}^{\mac} = \delta\gamma_{il} \sigma^{\mac}_{il} + \delta\kappa_{ij} \mu^{\mac}_{ij} = \left(\alpha^{(1)}_{il} + \levicivita_{jkl} x^{\mac}_k \beta_{ij}^{(1)}\right)  \sigma^{\mac}_{il} + \beta_{ij}^{(1)} \mu^{\mac}_{ij}
\end{align}
The equivalence between virtual work of internal forces at macroscale and microscale and arbitrariness of virtual parameters yields 
\begin{subequations}
\begin{align}
&\alpha_{ij}^{(1)}\rightarrow &  \sigma^{\mac}_{ij} &\approx \begin{cases} \dfrac{1}{V}\displaystyle\sum_e A_e l_e e^{Ne}_i t^e_j  & \text{using internal actions}\\[2mm]
\dfrac{1}{V}\displaystyle\sum_I x^I_i F^I_j & \text{using external actions}
\end{cases}\\
&\beta_{ij}^{(1)}\rightarrow &  \mu_{ij}^{\mathrm{mac}} &\approx \begin{cases} -\levicivita_{jkl}x^{\mac}_k \sigma^{\mac}_{il} + \dfrac{1}{V}\displaystyle\sum_e A_e l_e e^{Ne}_i \left(m^e_j + \levicivita_{jkl} x^{ce}_k t^e_l \right)  & \text{using internal actions}\\[2mm]
-\levicivita_{jkl}x^{\mac}_k \sigma^{\mac}_{il} + \dfrac{1}{V}\displaystyle\sum_I   x^I_i \left(Z^I_j+\levicivita_{jkl} x^I_k F^I_l  \right)& \text{using external actions}
\end{cases}
\end{align}
\end{subequations}
The equations are identical to Eqs.~\eqref{eq:fabric_stress_ext}, \eqref{eq:fabric_couple_stress_ext}, \eqref{eq:fabric_stress_int}, and \eqref{eq:fabric_couple_stress_int} derived by discretization of continuous expressions with neglected boundary-radius gap term.

The very same approach can be used for flux homogenization. The virtual pressure is adapted from Eq.~\eqref{eq:virt_press}, $\delta p = x_i \pi^{(1)}_i$, the first term is removed since, as discussed below Eq.~\eqref{eq:mmicrobalances_flux}, it leads to global balance of external actions which is assumed valid for our microscale system.The virtual pressure gradient, $\delta g$, is computed from pressures at nodes $I$ and $J$ 
\begin{align}
\delta g = \frac{\delta p_J - \delta p_I}{l} = \pi_i^{(1)} e^{\lambda}_i \label{eq:discrete_kinematics_transport}
\end{align}
The virtual power of internal actions for single conduit element $e$ connecting nodes $I$ and $J$ reads
\begin{align}
\delta W_{\intx}^e = S h j \delta g =  S h j  e^{\lambda}_i \pi_i^{(1)}
\end{align}
where $j$ is flux scalar. The average density of the virtual power of internal actions in the  discrete system is summation of the previous equation over all the conduit elements divided by volume. The average density of the virtual power of external actions is straightforward
\begin{align}
\delta \bar{w}_{\intx}^{\mic} &= \frac{1}{V} \sum_d S_d h_d j_d  e^{\lambda d}_i \pi_i^{(1)} & 
\delta \bar{w}_{\ext}^{\mic} &=-\frac{1}{V} \sum_P Q_P x^P_i \pi^{(1)}_i 
\end{align}
The macroscopic density of internal power reads
\begin{align}
\delta w_{\intx}^{\mac} = a_i \nabla_i \delta p  = \pi_i^{(1)} a_i
\end{align}
The equivalence of virtual power at micro and macroscale and arbitrariness of virtual parameter $\pi_i^{(1)}$ provides
\begin{align}
&\pi_{i}^{(1)}\rightarrow &  a^{\mac}_{i} &\approx \begin{cases} \dfrac{1}{V}\displaystyle\sum_d S_d h_d e_i^{\lambda d} j_d  & \text{using internal actions}\\[2mm]
-\dfrac{1}{V}\displaystyle\sum_P x^P_i Q_P & \text{using external actions}
\end{cases}
\end{align}
The result is identical to the formulas~\eqref{eq:fabric_flux_ext} and \eqref{eq:fabric_flux_int} derived by discretization of the continuous expressions with neglected boundary-radius gap term.

\section{Two dimensional version of the couple stress \label{sec:2D}}

Sometimes the discrete systems are formulated in two dimensions (2D) only. Although one can easily use the expressions derived for three-dimensional system, large portion of tensor components will be zero. It is therefore better to use two dimensional versions of the variables directly. The expressions for the stress tensor and flux vector remain the same, only the size of the involved tensors is reduced (e.g., two dimensional stress tensor has size $2\times 2$ while three dimensional version is $3\times 3$). A~fundamental difference appears in the expression for the couple stress, which becomes vector of size 2 in 2D contrary to a~second order tensor of size $3\times 3$ in 3D. The two components $\mu^{\text{(2D)}}_1$ and $\mu^{\text{(2D)}}_2$ of the two dimensional version correspond to components $\mu^{\text{(3D)}}_{13}$ and $\mu^{\text{(3D)}}_{23}$ of the three-dimensional couple stress. Also the Levi-Civita permutation tensor becomes the second order tensor, $\levicivita^{\text{(2D)}}_{jk}$, in two dimensions corresponding to the slice of the three-dimensional counterpart,  $\levicivita^{\text{(2D)}}_{jk}= \levicivita^{\text{(3D)}}_{3jk}$. External couples and internal couple tractions are scalars, the third components of the corresponding vectors in 3D. 

Using the three dimensional notation, expressions~\eqref{eq:fabric_couple_stress_ext} and \eqref{eq:fabric_couple_stress_int} read
\begin{align}
\upmu^{\mac}_{i3} &= \levicivita_{3jk} \sigma^{\mac}_{ij} x^{\mac}_k  + \frac{1}{V} \sum_I x^I_i \left(Z^I_3 + \levicivita_{3jk} x^I_j F^I_k\right) & 	\upmu^{\mac}_{i3} &= \levicivita_{3jk} \sigma^{\mac}_{ij} x^{\mac}_k  + \frac{1}{V} \sum_e A_e l_e e^{Ne}_i \left(m^e_3 + \levicivita_{3jk}x^{ce}_j t^e_k\right)
\end{align}	
The above equations in 2D notation become
\begin{align}
\boldmu^{\mac} &= \begin{cases} \boldsigma^{\mac} \otimes \x^{\mac}:\llevicivita^{\text{(2D)}} + \dfrac{1}{V} \displaystyle\sum_I \x_I \left[Z_I + \llevicivita^{\text{\text{(2D)}}}:\left(\x_I\otimes \boldF_I\right)\right] & \mathrm{using\ external\ actions} \\ \boldsigma^{\mac} \otimes \x^{\mac}:\llevicivita^{\text{(2D)}} + \dfrac{1}{V} \displaystyle\sum_e A_e l_e \e_{Ne} \left[m_e + \llevicivita^{\text{(2D)}}:\left(\x_{ce} \otimes \boldt_e\right)\right]  & \mathrm{using\ internal\ actions}
\end{cases}
\end{align}	

\begin{figure}[!tb]
\centering \includegraphics[width=0.8\textwidth]{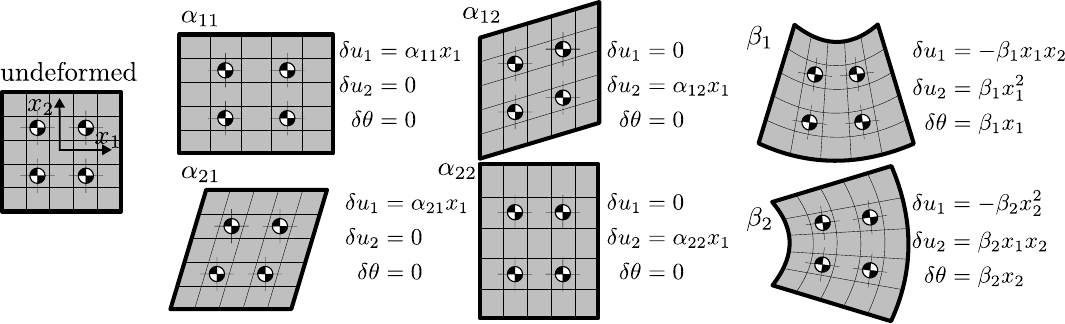}
\caption{Components of virtual displacement fields in 2D: $\alpha$ modes account for stretching and shearing with zero rotations, $\beta$ modes describe local rotations and curvatures.}\label{fig:2D_modes}
\end{figure}

It it worth showing the virtual kinematic modes that directly lead to these equations. The three-dimensional virtual kinematics~\eqref{eq:restricted_virt_kinematics} reduces to 
\begin{align}
\delta u_i & =x_j \alpha^{(1)}_{ji} + \levicivita_{ij} x_j x_k \beta_{k}^{(1)} & \delta \theta &=  x_j \beta_{j}^{(1)} \label{eq:constr_virt_kinematics_2D}
\end{align}
and Figure~\ref{fig:3D_modes} becomes Figure~\ref{fig:2D_modes}.

\section{Verification of equations for Poisson's problem \label{sec:verif_poisson}}
The verification starts with Poisson's problem in this section and continues with mechanics in the next one. All the discrete and finite element simulations presented here were computed in an~in-house open source code OAS (Open Academic Solver\footnote{\url{https://gitlab.com/kelidas/OAS}}). 

A~simple two-dimensional discrete structure generated by randomly placing circular particles into a~domain without overlapping is created. The diameters follow the Fuller curve with maximum diameter $d_{\max}=10$\,mm. All circles with diameter below 4\,mm were omitted (assuming their effect is homogenized and phenomenologically incorporated into the constitutive behavior). The power tessellation~\parencite{Eli17} is employed to generate convex polygonal shapes of particles, which continuously fill the space. 

Equation~\eqref{eq:discrete_kinematics_transport} is used as the discrete version of the kinematic equation, only the virtual variables are replaced by the real ones. The constitutive equation reads $j = -\lambda g$,
where flux $j$ and pressure gradient $g$ are scalars (analogies to the vectors $\bolda$ and $\boldg$ in the continuum). The permeability coefficient $\lambda$ is a~material parameter chosen here for the sake of simplicity as $\lambda=1$ (units are ignored in this section). Finally the balance equation states that the total mass flowing in and out of some control volume of volume $V$ associated with node $P$ must be equal:
\begin{align}
V_P c\, \dot{p}_P + \sum_{d\in P} S_d j_d  = V_P q_P  \label{eq:discrete_balance_transport}
\end{align}
where $c$ is capacity and $\dot{p}$ is time derivative of pressure. The form of the balance equation is the same for other physical problems but the terms might have different names. The capacity term will be omitted in the following analysis, only the steady state solution is explored. However the transient term can be moved to the right-hand side during evaluation of the macroscopic flux and treated as an~additional source term. This is done in the mechanical model in Sec.~\ref{sec:transient_verification} and it would work identically also here. 

The discrete model is generated within a~square domain of size 1$\times$1\,m$^2$, there are approx. 72\,000 conduit elements interconnecting approx. 48\,000 nodes. First a~standard patch test with constant flux is performed. The pressure, source term and flux in the domain read
\begin{align}
p &= 2\left(2x-1\right) + 2\left(2y-1\right) &  q &= 0 & \bolda &= \left(\begin{array}{c} -4 \\ -4\end{array} \right)
\end{align} 
The pressure is directly prescribed to all boundary nodes as Dirichlet boundary condition; the pressures at internal nodes are free variables. The discrete model in the present form (homogeneous material) satisfies the patch test and provides the exact values of the pressure in all the nodes (except for the errors due to machine precision). The proof is simple, one needs to show that the discrete balance equation is satisfied for every simplex control volume, $P$, for the exact solution. The flux given by kinematic and constitutive equations reads $j = \bolda\cdot\e_{\lambda} = -4\left(e^{\lambda}_1+e^{\lambda}_2\right)$. Since the Power tessellation ensures the faces are perpendicular to the vectors connecting the nodes $P$ and $Q$, the vector $\e_{\lambda}$ is an~outward normal vector for the whole simplex boundary, $\Gamma_P$. One can rewrite the balance equation as an~integration of the average flux over the enclosed control volume boundary and use the divergence theorem to prove that the balance equation
\begin{align}
\sum_{d\in P} S_d \bolda \cdot\e_{\lambda d}  = \bolda\cdot\int_{\Gamma_P} \e_{\lambda} \dd{\Gamma} = 0
\end{align}  
 is satisfied. The proof requires the faces to be perpendicular to the  $\vec{PQ}$ vectors and the control volume boundary must form an~enclosed surface. It is valid also in three dimensions. The mathematical steps are elucidated in detail in Refs.~\parencite[eq. 59]{Eli20} and \parencite{ZhaEli-24}.

Similar reasoning could be directly used to show that the macroscopic flux expressed by Eq.~\eqref{eq:fabric_flux_ext_exact} provides the exact macroscopic flux. One can take any connected set of nodes and create one aggregated control volume, $\tilde\Omega$, with enclosed surface, $\tilde\Gamma$, and normal surface vectors $\e_{\lambda}$. The external forces are now fluxes in conduit elements crossing the boundary. They can be smeared over the perpendicular face to transform the summation into integration. The macroscopic flux for any control volume therefore yields
\begin{align}
\bolda^{\mac} = \frac{1}{V} \sum_{d\in\tilde{\Gamma}} \x_{cd} \,\bolda \cdot \e_{\lambda d} = \frac{1}{V} \bolda \cdot
\int_{\tilde{\Gamma}}  \e_{\lambda}\otimes \x \dd{\tilde{\Gamma}} = \bolda\cdot\mathbf{1} = \bolda
\label{eq:flux_exact}
\end{align} 
For more details see Ref.~\parencite[eq. 60]{Eli20}.

\begin{figure}[!tb]
\centering \includegraphics[width=\textwidth]{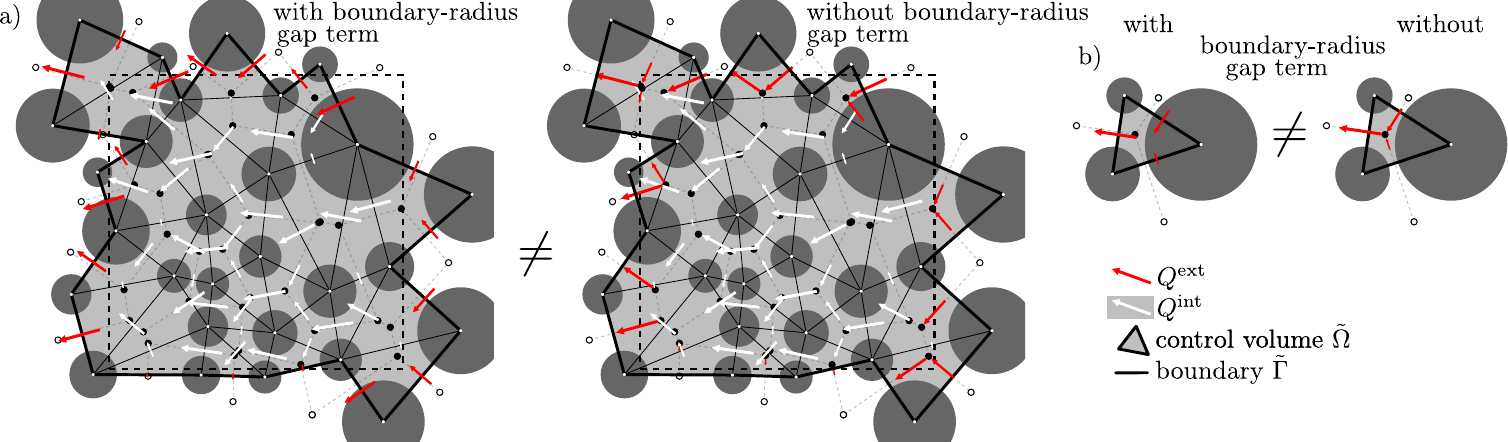}
\caption{Schematic description of two variants (with or without boundary-radius gap term) for evaluation of macroscopic flux in the discrete system: a) the control volume is dictated by a~square bin depicted by dashed line, b) the control volume is a~single triangle of weighted Delaunay triangulation.} \label{fig:bin_processed}
\end{figure}

The homogeneous value of macroscopic flux, as stated in Eq.~\eqref{eq:flux_exact}, is obtained only from Eq.~\eqref{eq:fabric_flux_ext_exact}, i.e., with the boundary-radius gap term. Neglecting it as done for Eqs.~\eqref{eq:fabric_flux_ext} and \eqref{eq:fabric_flux_int}
will result in deviations from the homogeneous value. Two variants are explored now, the first one with the boundary-radius gap term, and the second one without it. The second variant can be computed either from internal or external actions, the first variant can be evaluated directly from external actions or as the second variant with boundary-radius gap term added. The situation is sketched in Fig.~\ref{fig:bin_processed} to illustrate the distinction between these two variants.

\begin{figure}[!tb]
\centering \includegraphics[width=15cm]{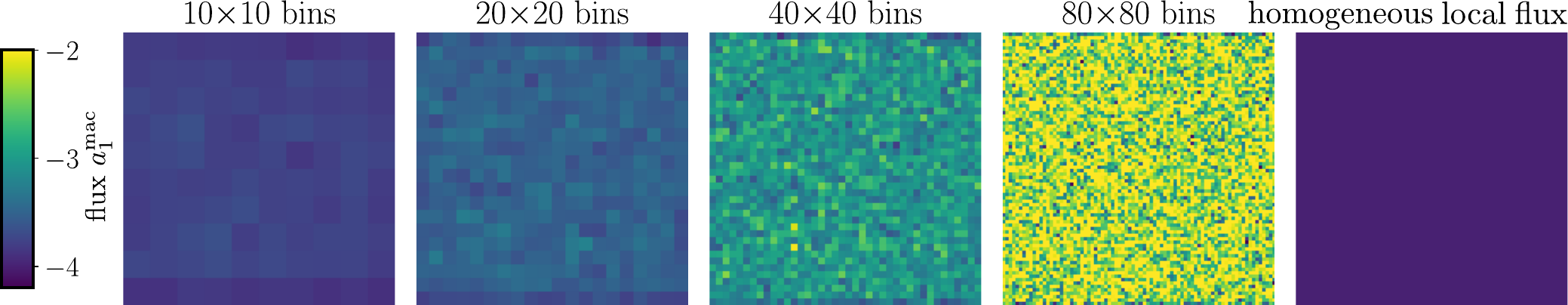}
\caption{Macroscopic flux component $a^{\mac}_1$ computed by the variant II (without boundary-radius gap term) using square bins of various sizes. Variant I (with boundary-radius gap term) always provides the exact solution. Fluxes $a^{\mac}_2$ have the same character. There are approximately 480, 120, 30, and 8 nodes in each bin, respectively.}\label{fig:poisson_maps_A}
\end{figure}

Let us now compute the macroscopic flux for both variants. The domain is divided into square bins; the control volume $\tilde{\Omega}$ encompasses all the nodes within each bin, see Fig.~\ref{fig:bin_processed}. Variant I always provides the correct homogeneous value as has been proven before. Results in $x$ direction from variant II are shown in Fig.~\eqref{fig:poisson_maps_A}; they depend on number of bins. The larger the bin relatively to the size of particles, the less the deviation from the correct solution. The simulation intentionally employed the dual network of power tessellation, as reference points can sometimes be distant from the centroids of triangular control volumes $\Omega_P$, and in some cases even outside the triangle. The difference between variants I and II, the boundary-radius gap term, is therefore accentuated. For mechanical problems, the reference points are typically better positioned and the difference is less severe. On average, the flux decreases as the number of bins increases. This is simply a~consequence of the source term position: when moved inward to the reference point, the flux appears in smaller volume but the total volume remains. In an~extreme of a~single point within the control volume (Fig.~\ref{fig:bin_processed} on the right-hand side), variant I (with boundary-radius gap term) still provides exactly the correct fluxes, but variant II (without boundary-radius gap term) results in zero flux because all the sources act at the same point and cancels each other.

\begin{figure}[!tb]
\centering \includegraphics[width=15cm]{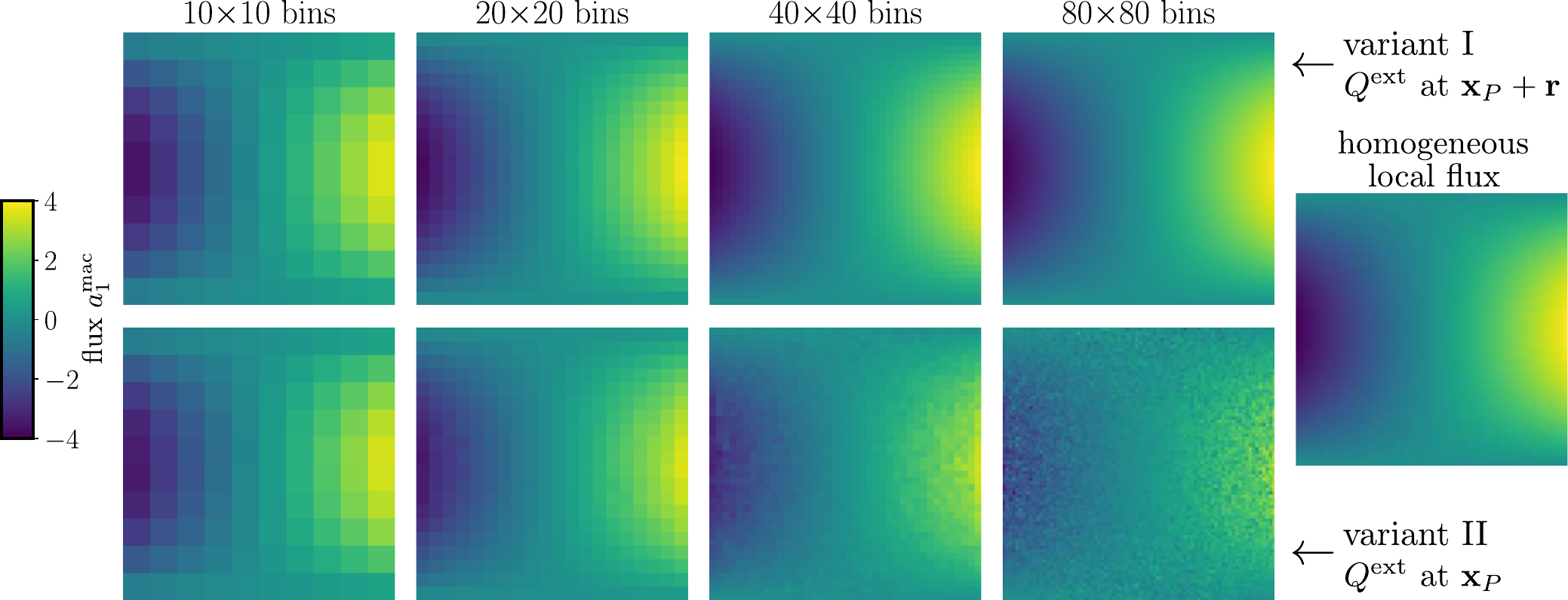}
\caption{Macroscopic flux component $a^{\mac}_1$ computed by the variants I \& II using square bins of various sizes. Fluxes $a^{\mac}_2$ have the same character.}\label{fig:poisson_maps_B}
\end{figure}

The second example for the validation of the macroscopic flux equations is taken from Ref.~\parencite{ManRus-14}. It uses the same geometry of $1\times 1$\,m$^2$. However, this time, the exact solution, load, and flux are defined as follows
\begin{align}
p &= 16x_1 x_2 \left(1-x_1\right)\left(1-x_2\right) &  q &= 32\left(x_1-x_1^2+x_2-x_2^2\right) & \bolda &= \left(\begin{array}{c}  16x_2\left(2x_1-1\right)\left(1-x_2\right) \\ 16x_1\left(2x_2-1\right)\left(1-x_1\right) \end{array} \right) \label{eq:patch2}
\end{align} 
Dirichlet boundary conditions were applied along the entire domain boundary, and the source was distributed to all the nodes in the domain according to their associated volume. This time, the discrete system does not provide the expected homogeneous solution from Eq.~\eqref{eq:patch2} because the discrete model cannot represent nonlinear pressure field exactly. Fluxes $a^{\mac}_1$ are evaluated with and without boundary-radius gap term and shown in Fig.~\ref{fig:poisson_maps_B} along with the homogeneous local fluxes. One can observe that for larger bins, both variants provide similar results. As the bin size decreases, the first variant maintains its quality but the second variant tends to push the results slightly towards zero. As argued in the previous example, the relative volume of the boundary layer with no flux increases with decreasing control volume, hence the macroscopic fluxes shift towards zero.

\section{Verification of mechanical equations}
The mechanical equations are not verified with a~help of the patch test, because the discrete structure does not provide solutions exactly corresponding to the homogenized continuous one with the only exception when Poisson's ratio is 0 (so called elastically uniform lattice, see Refs.~\parencite{Eli20}) or by introducing special techniques such as decoupling the normal strain component into the volumetric and deviatoric part~\parencite{CusRez-17} or adding auxiliary eigen-strains~\parencite{AsaIto-15,AsaAoy-17}. Instead, a~``real-world'' example of a~simple cantilever is analyzed. 

The kinematic equation providing strain and curvature was already presented as Eq.~\eqref{eq:virtual_discrete_kinematic_equation}. The virtual variables are now replaced by the real ones and the whole equation is transformed into two-dimensions. Also this time a~projection to the local coordinate system given by the normal direction $\e_N$ and tangential direction $\e_M$ is added. Index $\alpha\in\left\{ N, M \right\}$ will be used to distinguish between directions in the local coordinate system hereinafter.
\begin{align}
\epsilon_{\alpha} &= \frac{1}{l}\left[\boldu_J-\boldu_I - \llevicivita^{\text{(2D)}}\cdot(\theta_J \boldc_J - \theta_I \boldc_I) \right]\cdot\e_{\alpha} &
\kappa &= \frac{1}{l}\left(\theta^J - \theta^I \right) \label{eq:discrete_kinematic_equation}
\end{align}
Note the different sign in front of the Levi-Civita permutation symbol in the 2D version due to switch of its indices to allow contraction with vector $\boldc$.  

The projection of strains into the local reference system enables to use different stiffness in the normal direction ($E_0$) and tangential directions ($\alpha E_0$). This ability is critical for adjusting Poisson's ratio of the system~\parencite{KuhDadd-00,Eli20}. The constitutive equation reads
\begin{align}
t_N & =E_0 \epsilon_N & t_M &= \alpha E_0 \epsilon_N & m & = \beta  \frac{E_0 I}{A} \kappa = \beta \frac{E_0 A^2}{12} \kappa \label{eq:discrete_constitutive}
\end{align}
$A$ denotes area of the contact and $I$ denotes its moment of inertia. In two dimensions the area becomes length of a~line with units $[m]$ and moment of inertia is then $A^3/12$, hence the second equality in the last equations. A complete derivation of the last equation is presented in appendix~\ref{app:beta}. There are three elastic parameters introduced in the constitutive equation. Two of them, the normal modulus $E_0$ and the ratio between the tangential and normal moduli $\alpha$, are well established in the literature~\parencite{CarBaz97,CusBaz-06}. The third parameter, $\beta$, is added to control the bending stiffness. When $\beta=1$ the resulting moment acting between particles $I$ and $J$ exactly corresponds to the continuous integration of the moments caused by tractions $\boldt$ on eccentricities  over the facet area, see appendix~\ref{app:beta}. For the present verification, elastic parameters are $E_0=40$\,GPa and $\alpha=0.3$ while $\beta$ parameter varies between 0 and $10^5$. The large values of $\beta$ are included to explore asymptotic behavior of the model, they do not correspond to any real microstructures or materials..

The last governing equation is the balance of linear and angular momentum in two dimensions
\begin{subequations} \label{eq:discrete_balance_eq}
\begin{align}
\sum_{e\in I} A_e t^e_{\alpha} \e_{\alpha e}  & = \rho V_I \ddot{\boldu}_I + \boldM_{u\theta} \ddot{\theta}_I - V_I \boldF_I
\\ 
\sum_{e\in I} A_e \left[ m_e + t^e_{\alpha}\llevicivita^{\text{(2D)}}:\left(\boldc_{Ie}\otimes \e_{\alpha e} \right)\right]  & = M_{\theta}\ddot{\theta}_I + \boldM_{\theta u}\cdot \ddot{\boldu}_I - V_I Z_I
\end{align}
\end{subequations}
where the material density was set to 2400\,kg/m$^{3}$. Tensors $\boldM$ denote various inertia terms calculated according to equations in Ref.~\parencite{EliCus22}.

The cantilever generated for verification of mechanical equations has a~depth of $D=3$\,m and a~length of $S=12$\,m. The large dimensions were chosen to have a~large number of particles when evaluating macroscopic quantities. It is fixed ($u_1=u_2=\theta=0$) at the left-hand side and loaded with a~force of $P=100$\,kN at the right-hand side. The force is distributed along the entire right-hand side boundary using a~linear constraint. The vertical displacements of nodes on the right-hand side are constrained to follow the vertical displacement of the central auxiliary node. The same discretization using particles generated by the Fuller curve with maximum radius $d_{\max}=10$\,mm and power tessellation is used. In total the model has 1\,034\,196 degrees of freedom.

\begin{figure}[!tb]
	\centering \includegraphics[width=0.6\textwidth]{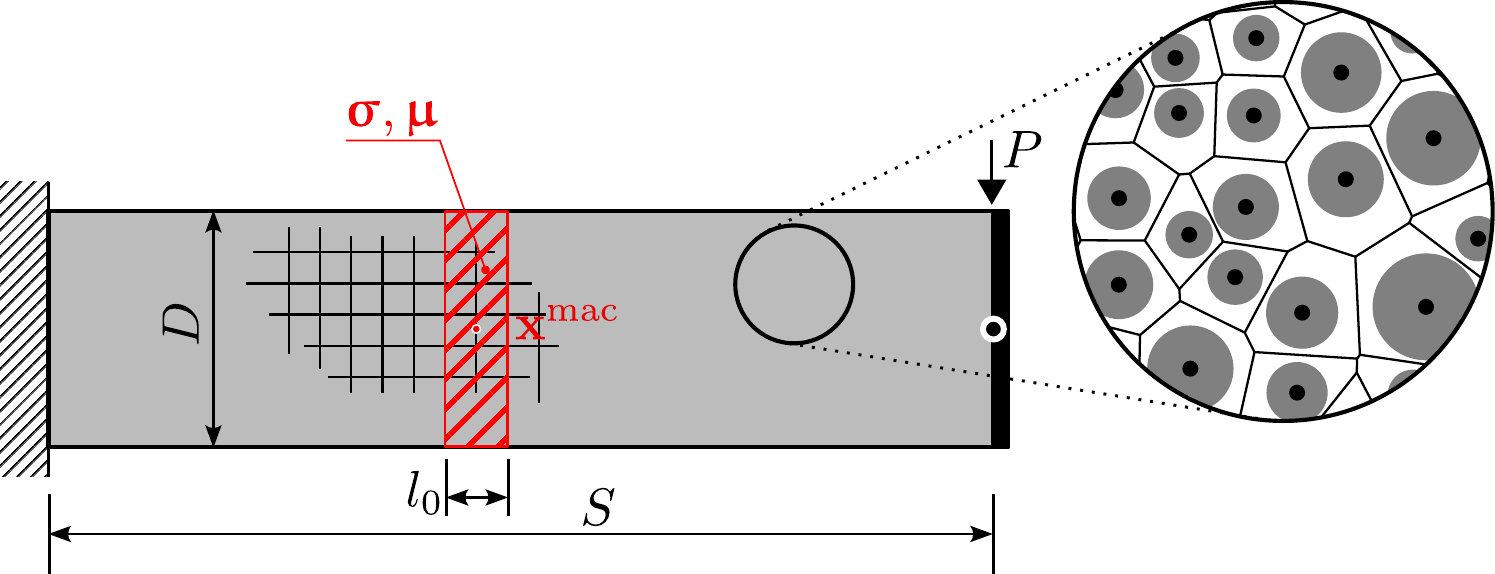}
	\caption{Model of a~cantilever used in both steady state and transient examples.}\label{fig:model}
\end{figure}

It is beneficial to evaluate the approximate macroscopic behavior of the material for comparison. For this purpose, a~periodic unit cell with periodic boundary conditions is loaded with unit strain and curvature in all directions, following the homogenization theory developed by~\textcite{ForPra-01,EliCus22}. Relatively large RVE size $0.2\times 0.2$\,m$^2$ is used, even though the mean results are independent on this size. The advantage of larger RVE is its better statistical representativeness, it decreases variability of the results with a~change of internal structure. The resulting loading stresses were collected into a~homogenized tensor of elastic constants that approximately represents the mechanical behavior of the material. This tensor is directly used in a~continuum model of a~Cosserat solid, hereinafter referred to as the \emph{homogenized} model. The domain for the \emph{homogenized} model is discretized into bilinear (both displacement and rotations) square Cosserat finite elements~\parencite{ZhaWan-05}, each of the same size $0.2\times0.2$\,m$^2$, and is solved using the finite element method.

\begin{table}[!bt]
\begin{center}
\begin{tabular}{ccccccc}
\Xhline{3\arrayrulewidth}
$\beta$ & $\lambda$ & $\mu$ & $\mu_c$ & $\ell$ & $E$ & $\nu$ \\ 
{[-]} 	& [GPa] & [GPa] & [GPa] 	& [cm] & [GPa] & [-] \\ \hline
0		&8.58	&11.42	&6.0	&4.55	&27.74	&0.214\\
1		&8.28	&11.723	&6.0	&4.55	&28.3	&0.207\\
10		&8.06	&11.94	&6.0	&4.60	&28.7	&0.201\\
10$^2$	&7.99	&12.01	&6.0	&5.03	&28.8   &0.200\\
10$^3$	&7.98	&12.02	&6.0	&8.18   &28.8	&0.200\\
10$^4$	&7.98	&12.02	&6.0	&22.0	&28.8	&0.200\\
10$^5$	&7.98	&12.02	&6.0	&68.2	&28.8	&0.200\\\Xhline{3\arrayrulewidth}

\end{tabular}
\end{center}
\caption{Material parameters of plane strain Cosserat continuum material identified on a~periodic volume of size $0.2\times 0.2$\,m$^2$ for various values of $\beta$ parameter. \label{tab:CosseratParams}}
\end{table}

A~third model is included in our comparison, referred to as the \emph{simplified} Cosserat model. In this model, the tensors of elastic constants from the \emph{homogenized} model are approximated using a~simple Cosserat model consisting of four parameters~\parencite{ZhaWan-05}. These parameters include the Lamé parameters $\lambda$ and $\mu$, as well as additional parameters for Cosserat effects, $\mu_c$ and $\ell$. In the Voigt notation, with stresses ordered as $\sigma_{11}$, $\sigma_{22}$, $\sigma_{21}$, $\sigma_{12}$, $\mu_{1}$, and $\mu_{2}$, the tensor of elastic constants can be expressed as~\parencite{ZhaWan-05}
\begin{align}
\mathbf{D} = \left(\begin{array}{cccccc}
\lambda+2\mu & \lambda & 0 & 0 & 0 & 0 \\
\lambda & \lambda+2\mu & 0 & 0 & 0 & 0 \\
0 & 0 & \mu+\mu_c & \mu-\mu_c & 0 & 0 \\
0 & 0 & \mu-\mu_c & \mu+\mu_c & 0 & 0 \\
0 & 0 & 0 & 0 & 4\mu\ell^2 & 0 \\
0 & 0 & 0 & 0 & 0 & 4\mu\ell^2
\end{array} \right) \label{eq:CosseratTensor}
\end{align}
The four constitutive parameters were determined through a~least-square fitting of the material tensor of elastic constants using the simplified theory. These parameters effectively capture the influence of the bending stiffness $\beta$. Repeating the fitting process for different bending stiffness values, the characteristics listed in Table~\ref{tab:CosseratParams} are obtained. It is evident that the Lamé parameters change only marginally with varying $\beta$,  while parameter $\mu_c$ is constant, and the Cosserat characteristic length $\ell$ is significantly affected. The characteristic length increases from values close to 4.6\,mm (approx. the minimum aggregate diameter) to nearly 0.7\,m. For $\beta$ values below $10^3$, the Cosserat effects are negligible, and one can use the Cauchy continuum (for which $\boldtheta_m=\bm{0}$, there is no independent rotational degree of freedom and the stress tensor is symmetric). For added convenience, the table also includes the elastic modulus, $E$, and Poisson's ratio, $\nu$, which are related to the Lamé constants through the expressions $E = \mu(3\lambda+2\mu)/\left(\lambda+\mu\right)$ and $\nu=\lambda/\left[2(\lambda+\mu)\right]$.

As an example, tensors of elastic constants computed on RVE with $\beta=0$ and $\beta=10^5$ are shown in SI units bellow in Eq.~\eqref{eq:ela_tens_RVE} next to its optimal counterparts \eqref{eq:ela_tens_Cosserat} achieved by the simplified Cosserat model from Eq.~\eqref{eq:CosseratTensor}.
\begin{subequations}
\begin{align}
\stackrel{\mbox{homogenized RVE, $\beta=0$}}{\scriptsize
\left(\begin{array}{cccccc}
31.45 & 8.55 & -0.19 & -0.19 & 0.09 & -0.02\\
8.55 & 31.45 & 0.19 & 0.19 & -0.02 & -0.03\\
-0.19 & 0.19 & 17.31 & 5.31 & 0.00 & -0.01\\
-0.19 & 0.19 & 5.31 & 17.31 & -0.01 & -0.03\\
0.09 & -0.02 & 0.00 & -0.01 & 0.10 & 0.00\\
-0.02 & -0.03 & -0.01 & -0.03 & 0.00 & 0.09
\end{array} \right)}&\scriptsize\times 10^9
&
\stackrel{\mbox{homogenized RVE, $\beta=10^5$}}{\scriptsize
\left(\begin{array}{cccccc}
32.05 & 7.95 & -0.20 & -0.20 & 0.09 & -0.02\\
7.95 & 32.05 & 0.20 & 0.20 & -0.02 & -0.04\\
-0.20 & 0.20 & 17.90 & 5.90 & 0.01 & -0.01\\
-0.20 & 0.20 & 5.90 & 17.90 & -0.01 & -0.03\\
0.09 & -0.02 & 0.01 & -0.01 & 22.16 & 0.72\\
-0.02 & -0.04 & -0.01 & -0.03 & 0.72 & 22.49
\end{array} \right)}&\scriptsize\times 10^9
\label{eq:ela_tens_RVE}
\\
\stackrel{\mbox{Cosserat model, $\beta=0$}}{
\scriptsize
\left(\begin{array}{cccccc}
31.42 & 8.58 & 0 & 0 & 0 & 0\\
8.58 & 31.42 & 0 & 0 & 0 & 0\\
0 & 0 & 17.42 & 5.42 & 0 & 0\\
0 & 0 & 5.42 & 17.42 & 0 & 0\\
0 & 0 & 0 & 0 & 0.09 & 0\\
0 & 0 & 0 & 0 & 0 & 0.09
\end{array} \right)}&\scriptsize\times 10^9
&
\stackrel{\mbox{Cosserat model, $\beta=10^5$ }}{
\scriptsize
\left(\begin{array}{cccccc}
32.02 & 7.98 & 0 & 0 & 0 & 0\\
7.98 & 32.02 & 0 & 0 & 0 & 0\\
0 & 0 & 18.02 & 6.02 & 0 & 0\\
0 & 0 & 6.02 & 18.02 & 0 & 0\\
0 & 0 & 0 & 0 & 22.33 & 0\\
0 & 0 & 0 & 0 & 0 & 22.33
\end{array} \right)}&\scriptsize\times 10^9
\label{eq:ela_tens_Cosserat}
\end{align}
\end{subequations}

\subsection{Steady state}

\begin{figure}[!b]
\centering \includegraphics[width=\textwidth]{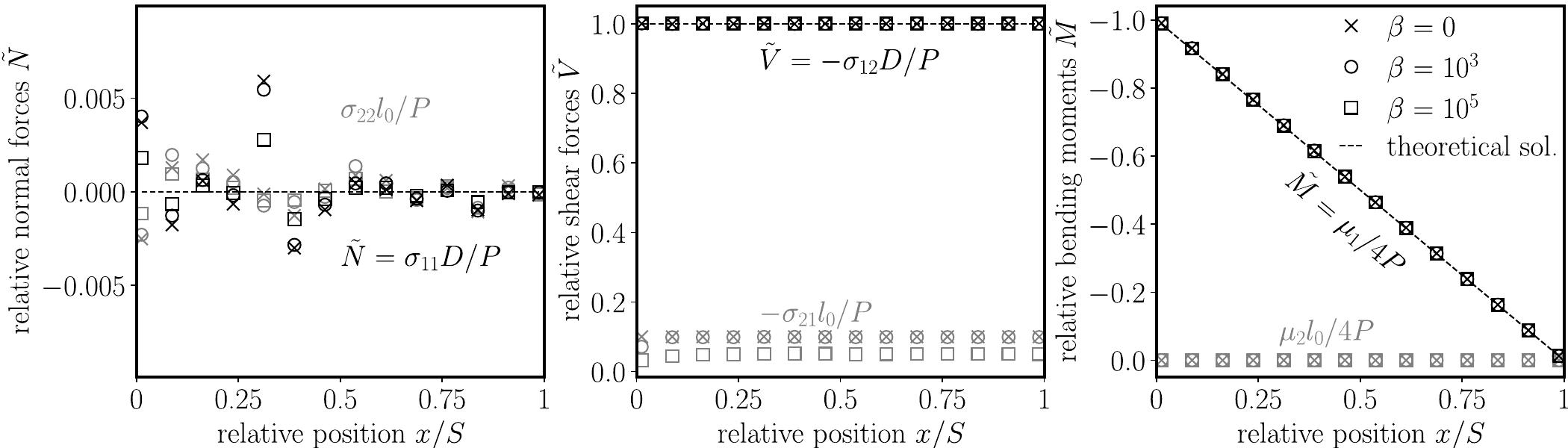}
\caption{Relative internal forces in the cantilever computed from discrete 2D solution.}\label{fig:statics_intfor}
\end{figure}

Firstly,  stress and couple stress components in vertical columns of width $\lRVE=D/10=0.3$\,m are computed with respect to the macroscopic point $\x^{\mac}$ located at the center of each column. The arrangement is illustrated in Fig.~\ref{fig:model} in red color. There are 40 such columns. The values of stresses and couple stresses are presented in Fig.~\ref{fig:statics_intfor} as relative internal forces. The normal and shear forces are derived from stresses $\sigma_{11}$ and $\sigma_{12}$, respectively, by multiplying them by the depth $D$ and normalizing them by the force $P$. The negative sign in the shear force arises from the opposite direction of shear stress and the standard convention of positive shear force. The bending moment is computed by taking the couple stress $\mu_1$, multiplying it by the depth $D$, and normalizing it by the maximum moment $16 P$. It's worth mentioning that these values precisely correspond to the theoretical values computed by static analysis of a~cantilever beam for all $\beta$ parameters, although only some of them are depicted. 

\begin{figure}[!b]
\centering \includegraphics[width=\textwidth]{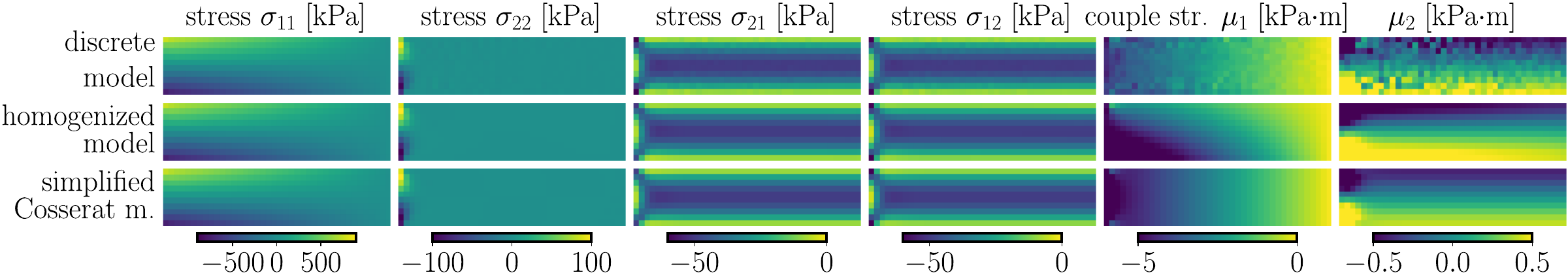}
\caption{Macroscopic stresses and couple stresses in discrete and continuous models for $\beta=0$.}\label{fig:statics_maps_A}
\end{figure}

\begin{figure}[!b]
\centering \includegraphics[width=\textwidth]{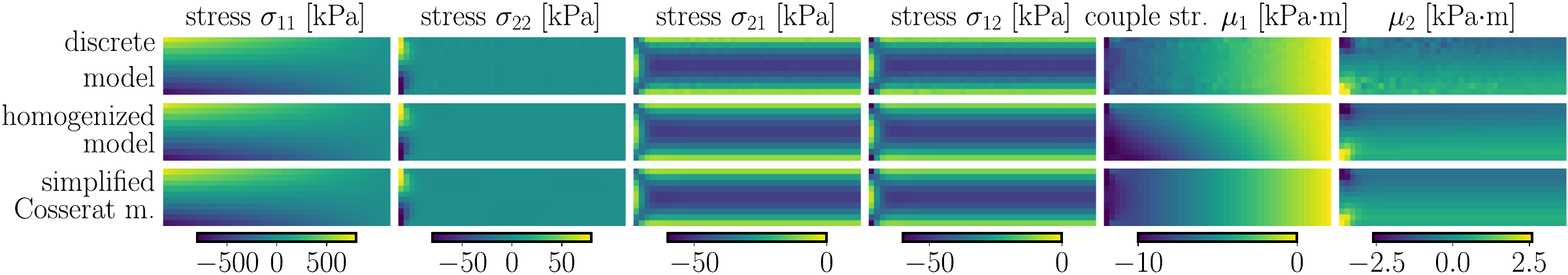}
\caption{Macroscopic stresses and couple stresses in discrete and continuous models for $\beta=10^3$.}\label{fig:statics_maps_B}
\end{figure}

\begin{figure}[!t]
\centering \includegraphics[width=\textwidth]{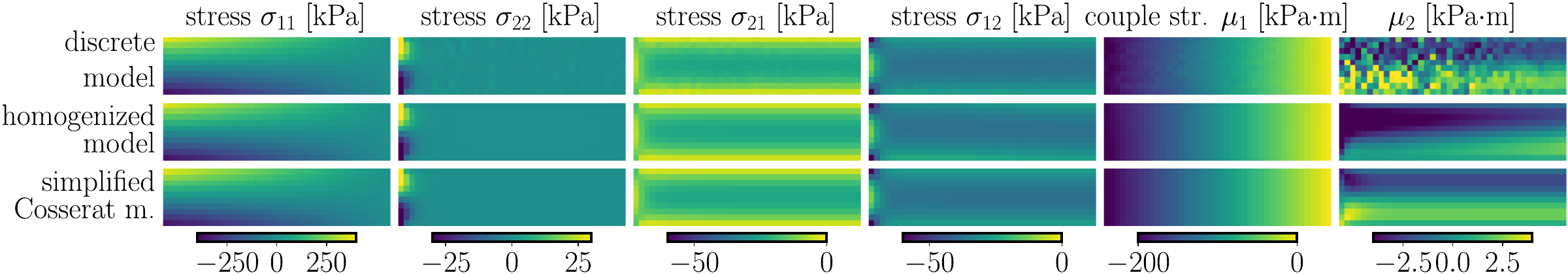}
\caption{Macroscopic stresses and couple stresses in discrete and continuous models for $\beta=10^5$.}\label{fig:statics_maps_C}
\end{figure}

Secondly, the steady-state solution of the mechanical model is computed for different $\beta$ parameters, considering all three model types: \emph{discrete}, \emph{homogenized}, and \emph{simplified} Cosserat models. Stress and couple stress values from integration points of the continuum models were averaged over integration points of four elements within bins of size $\lRVE\times\lRVE$ with respect to the bin central node, following equations~\eqref{eq:macrostress_internal} and \eqref{eq:macrocouplestress_internal}. The same process is applied to the \emph{discrete} model, averaging the values in bins of size $\lRVE \times \lRVE$ and employing the discrete equations \eqref{eq:fabric_stress_ext_exact} and \eqref{eq:fabric_couple_stress_ext_exact}, i.e., variant I with the boundary-radius gap, while always considering $\x^{\mac}$ at the bin center. Figures~\ref{fig:statics_maps_A}, \ref{fig:statics_maps_B}, and \ref{fig:statics_maps_C} demonstrate the reasonable correspondence between the models for $\beta$ values of $0$, $10^3$, and $10^5$, respectively. The omitted $\beta$ values exhibit similar correspondence.

The \emph{discrete} model aligns well with the \emph{homogenized} model. The observed differences can be attributed to the reduced kinematics of the continuum model as well as lack of statistical representativeness of the periodic unit cell. However, it is worth noting that there is also relatively good agreement with the \emph{simplified} Cosserat model, which also suffers from incomplete information about the material behavior. Note that as the bending stiffness increases through the parameter $\beta$, stress decreases while couple stress increases. This implies that the load is predominantly transferred by stress for low $\beta$ values and by couple stress for high $\beta$ values. The largest scatter is seen for couple stress $\mu_2$, which is relatively small (compared to $\mu_1$), and therefore largely affected by the heterogeneity. It is expected that the scatter would be reduced by using larger bins (the current bins have approximately 830 particles).

\subsection{Transient regime \label{sec:transient_verification}}
The same problem is analyzed in the transient regime for all three model types: \emph{discrete}, \emph{homogenized}, and \emph{simplified} Cosserat models, with varying $\beta$ parameters. The simulation begins in an~unloaded initial state. At time $0$\,s, a~constant load of $P=100$\,kN is applied. Time is discretized using the implicit generalized-$\alpha$ method~\parencite{ChuHul93} with a~spectral radius of 0.8 and time step of 0.15\,ms. In both continuum models the specific moment of inertia, $J_{\rho}$, from Eq.~\eqref{eq:continuum_angular_balance} was set to $d_{\max}^2/8$, which corresponds to a~specific moment of inertia of a~disc with radius~$d_{\max}$. However, the response is practically insensitive to this parameter.

\begin{figure}[!tb]
\centering \includegraphics[width=12cm]{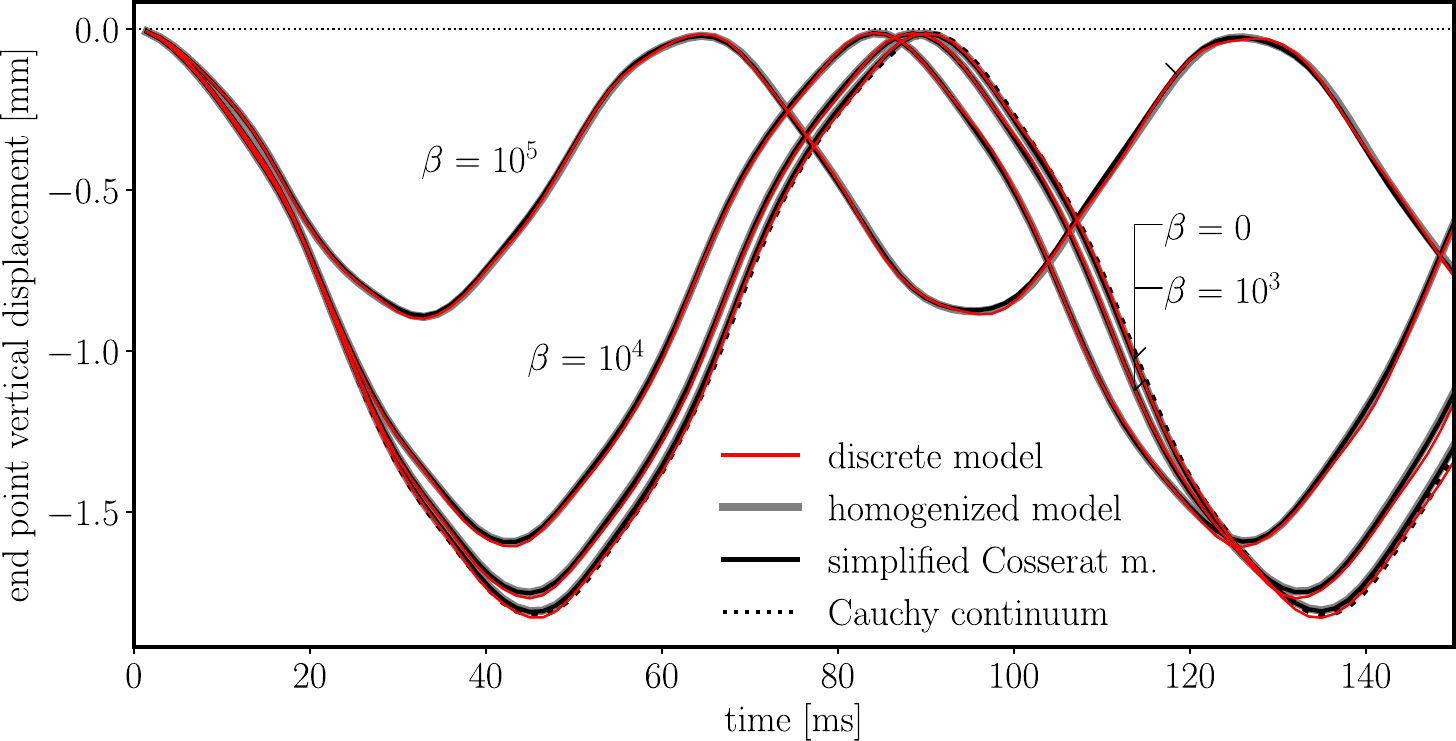}
\caption{Vertical displacements of the loaded point in time computed with various models of the cantilever depicted in Fig.~\ref{fig:model}.}\label{fig:dynamics_response}
\end{figure}

The vertical displacements of the node where load $P$ is applied are shown in Fig.~\ref{fig:dynamics_response} for all three models and selected values of $\beta$. An~additional model, the \emph{Cauchy} continuum, is included. This model represents a~standard continuous homogeneous isotropic model with symmetric shear stresses, using elastic parameters from the first row of Tab.~\ref{tab:CosseratParams}, $E=30.2$\,GPa and $\nu=0.18$. One can observe excellent agreement between different models for all values of $\beta$. The \emph{homogenized} model's response closely matches the \emph{discrete} model's response, with differences arising from the reduced kinematic description of the finite element approximation and different internal heterogeneity of the discrete model within the periodic cell and the full model. The \emph{simplified} Cosserat model shows also negligible deviations despite, additionally, it has incomplete information about the material behavior.The \emph{Cauchy} model closely resembles the remaining three models for $\beta=0$, providing further evidence that the Cosserat effects are negligible in this case.

\begin{figure}[!tb]
\centering \includegraphics[width=14cm]{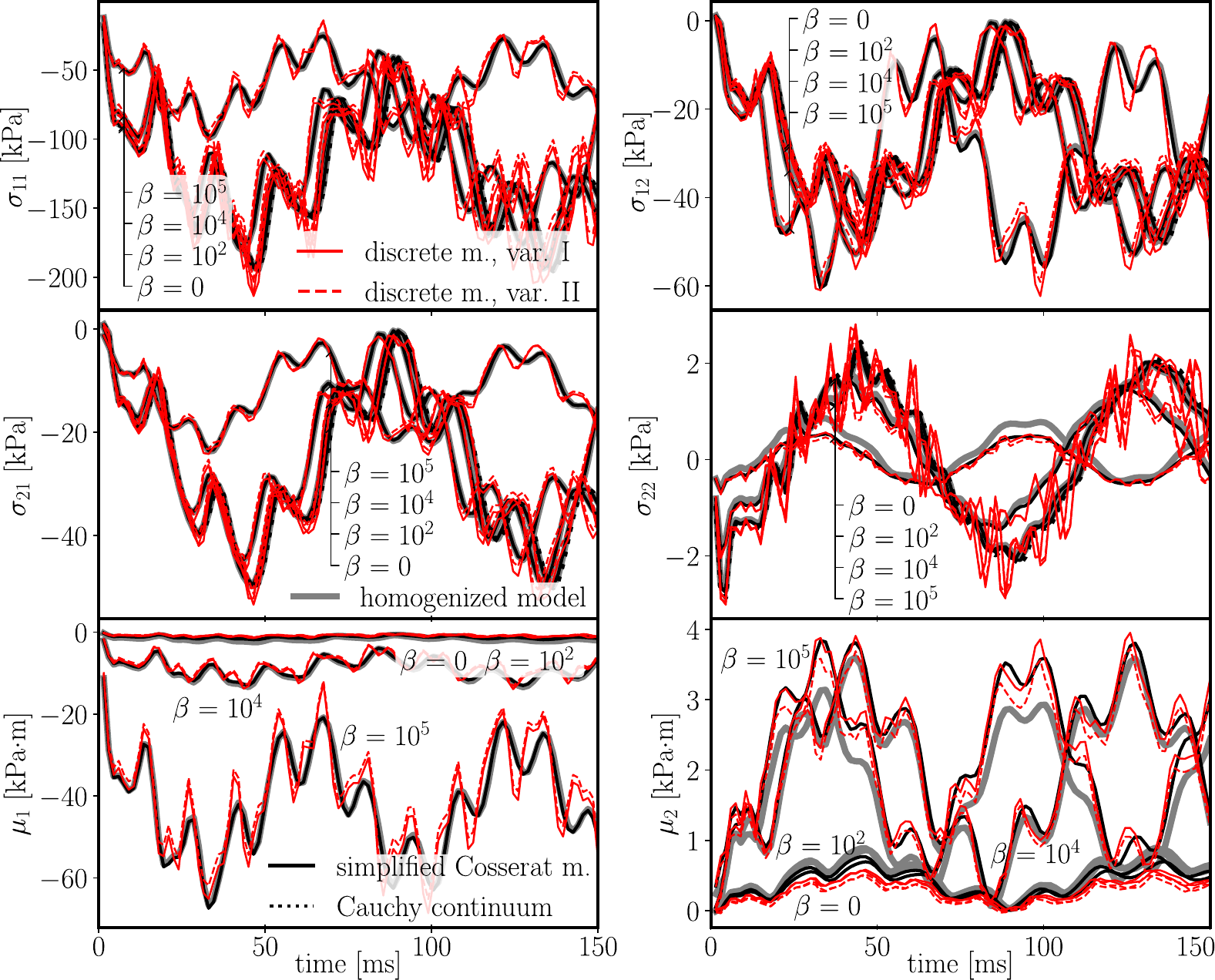}
\caption{Macroscopic stress and couple stress components evaluated for a~single bin from results computed with various models of the cantilever depicted in Fig.~\ref{fig:model}.}\label{fig:dynamics_stress}
\end{figure}

Finally, stress and couple stress tensors are computed over a~single bin of size $\lRVE\times\lRVE$ and their evolution in time is plotted in Fig.~\ref{fig:dynamics_stress}. For the \emph{discrete} model, macroscopic quantities are evaluated in variant I (with the boundary-radius gap  term) and variant II (without it) and $\x^{\mac}$ at the bin center. One can see that the number of particles is already sufficient for the two variants being almost indistinguishable. In the continuum models, macroscopic quantities are evaluated using 16 integration points within four involvedfinite elements, following Eqs.~\eqref{eq:macrostress_internal} and \eqref{eq:macrocouplestress_internal}. The selected bin for comparison is located in the 31$^{\text{st}}$ column (out of 40) and 2$^{\text{nd}}$ row (out of 10), measured from the bottom-left corner. This selection is arbitrary. Only four $\beta$ values are selected, but the agreement is similar also for the remaining $\beta$ values. One can once more see diminishing stresses and increasing couple stresses as $\beta$ grows. For $\beta=0$ the results from discrete and Cosserat models match the data from \emph{Cauchy} model.

\section{Conclusions}
Two sets of expressions have been derived to evaluate the macroscopic flux vector in Poisson's problems, as well as the macroscopic stress tensor and couple stress tensor in mechanical problems, for a~finite RVE of both continuous domains and discrete systems. The first set utilizes external forces, moments, and sources, while the second set uses internal tractions, couple tractions, and fluxes. The derivation demonstrates which virtual work balances between the micro and macroscale are satisfied and which are omitted in the construction of these equations. The choice of the micro-macro balances to satisfy is linked with the virtual deformation modes that are allowed at the microscale. The couple stress expression features a~new term that depends on the position of the macroscopic point into which the microstructure is lumped.

Extensive verification allows us to draw the following conclusions.
\begin{itemize}
\item All of the equations remain valid in both steady-state and transient regimes, providing the inertia forces and capacity terms are considered as additional external actions.
\item Equations based on internal actions omit the true location of external actions and, therefore, always introduce some error. The same is true when using external actions shifted into particle reference points. Both of these formulations can be corrected by summing the contributions of external actions acting at their true locations, so called boundary-radius gap term.
\item Discrete models with independent rotational degrees of freedom  macroscopically correspond to Cauchy continuum, unless a~large, unrealistic bending stiffness at the particle contacts is introduced. In such a~case, Cosserat continuum provides better macroscopic representation. These findings are in agreement with Refs.~\parencite{ForPra-01,CusRez-17}.
\item The asymptotic expansion homogenization of heterogeneous Cosserat media presented in Ref.~\cite{ForPra-01} studies the limit case of infinitely small representative volume. The work presented here considers a~finite size volume and, therefore, adds to the macroscopic couple stress also effect of micro-stresses on a~distance. These micro-stresses then cause couple stress at the macroscale even when it is of Cauchy type.
\end{itemize}

These derived results are directly applicable to heterogeneous discrete and continuous systems with nonlinear or coupled constitutive equations because these equations are not employed in the derivation. The verifications were conducted for a~specific discrete model based on power tessellation and weighted Delaunay triangulation, but the derived equations can be applied to different kinds of discrete models, including lattice models~\parencite{EddPun-20,ChaLia-23}, Lattice-Discrete Particle Model (LDPM)~\parencite{CusPel-11,MerPat-20,FasIch-22}, or various forms of Discrete Element Models (DEM)~\parencite{WanGao-20,KrzNit-23}, as long as they share fundamentally the same kinematic and balance equations.

\appendix
\section{Analytical integration of moment of traction \label{app:beta}}

This appendix proves that setting $\beta=1$ provides (in elastic regime and two dimensions) results exactly corresponding to the bending stiffness given by traction $\boldt$ on eccentricity $\boldc_I$. The discrete models typically integrate the moment of traction with very few integration points, in the most cases only a~single one. For $\beta=1$ the obtained results are identical to integrating these moments continuously.

\begin{figure}
\centering \includegraphics[width=3.6cm]{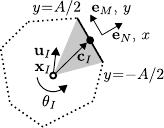}
\caption{Variables used in the integration of internal couple traction and moment of traction.} \label{fig:beta_effect}
\end{figure}

Let us formulate the local reference system as shown in Fig.~\ref{fig:beta_effect}. The $y$ coordinate runs along the facet, it is zero at its centroid. Traction at any position can be easily computed from the constitutive~\eqref{eq:discrete_constitutive} and kinematic~\eqref{eq:discrete_kinematic_equation} equations
\begin{align}
t_{\alpha}(y) &= \frac{E_{\alpha}}{l}\left[\boldu_J-\boldu_I  - \llevicivita\cdot\left(\theta_J \left(\boldc_J + y \e_M\right) - \theta_I \left(\boldc_I  + y \e_M\right) \right) \right]\cdot\e_{\alpha} = t_{\alpha}(0) - y\frac{E_{\alpha}}{l}\left(\theta_J  - \theta_I \right) \llevicivita : \left(\e_{\alpha} \otimes \e_{M}\right)
\end{align}
where $E_N = E_0$ and $E_M = E_0 \alpha$. Traction $\boldt(0)$ is the traction computed at the facet centroid at coordinate $y=0$. For the reference system given by orthonormal vectors $\e_N$ and $\e_M=(-e^N_2,\,e^N_1)$ it is easy to derive the following property  
\begin{align}
\llevicivita:(\e_{\alpha} \otimes \e_{\beta}) &= \begin{cases} 0 & \alpha=\beta \\ 1 & \alpha=N \text{ and } \beta=M \\ -1 & \alpha=M \text{ and } \beta=N \end{cases} \label{eq:cross_of_reference}
\end{align}
The traction along the facet therefore becomes 
\begin{align}
t_{\alpha}(y) &= t_{\alpha}(0) - y\frac{E_0}{l}\left(\theta_J  - \theta_I \right)\delta_{\alpha N}
\end{align}
where $\delta_{\alpha\beta}$ is the Kronecker's delta tensor. 

The internal couple applied to particle $I$ due to traction $t_{\alpha}(y)$ with arm  $\boldc_I + y \e_M$ is integrated over the facet. 
\begin{subequations}
\begin{align}
\int_{-A/2}^{A/2} t_{\alpha}(y) \llevicivita:\left[ \left( \boldc_I + y \e_M \right) \otimes \e_{\alpha}\right]   \dd{y} &=
\int_{-A/2}^{A/2} \left[t_{\alpha}(0) - y\frac{E_0}{l}\left(\theta_J  - \theta_I \right)\delta_{\alpha N}\right] \llevicivita:\left[ \left( \boldc_I + y \e_M \right) \otimes \e_{\alpha}\right]   \dd{y}\\
&= t_{\alpha}(0) \llevicivita:\left( \boldc_I \otimes \e_{\alpha}\right)A - \frac{E_0}{l}\left(\theta_J  - \theta_I \right) \delta_{\alpha N} \llevicivita:\left( \e_M \otimes \e_{\alpha}\right)  \frac{A^3}{12}\\
&= t_{\alpha}(0) \llevicivita:\left( \boldc_I \otimes \e_{\alpha}\right)A + \frac{E_0 A^3}{12l}\left(\theta_J  - \theta_I \right)
\end{align}
\end{subequations}
Equation~\eqref{eq:cross_of_reference} and $\int_{-A/2}^{A/2} y \dd{y}=0$ was used.	

Based on the integration of the moment at only one integration point at $y=0$ using the bending stiffness given by Eq.~\eqref{eq:discrete_constitutive}, the internal couple acting at particle $I$ reads
\begin{align}
t_{\alpha}(0) A \llevicivita:\left( \boldc_I \otimes \e_{\alpha}\right) + \beta\frac{E_0 A^2} {12 l} \left(\theta_J-\theta_I \right) A
\end{align}
This result is identical to the analytical integration above when $\beta=1$.
Of course this result is valid only in the elastic regime. If the constitutive model becomes nonlinear, the bending stiffness actually complicates the formulation as one needs inelastic constitutive law also for bending.

\section*{Acknowledgement} Jan Eliáš acknowledges financial support from the Czech Science Foundation under project number 23-04974S. Gianluca Cusatis acknowledges support by the Engineering Research and Development Center (ERDC) – Construction Engineering Research Laboratory (CERL) under Contract No. W9132T22C0015.  We also thank prof. Milan Jirásek from the Czech Technical University for a~discussion at the beginning of the work. 

\section*{Data availability}
The numerical simulations are computed using open source code OAS (Open Academic Solver) available at \url{https://gitlab.com/kelidas/OAS}, commit de14ee194f from 1.\,11.\,2024.  Input files for OAS as well as outputs and Python script used in this study are available at DOI \href{https://doi.org/10.5281/zenodo.14502116}{\texttt{10.5281/zenodo.14502116}}.

\printbibliography
\end{document}